\documentclass[11pt,a4paper]{JHEP3}
\pdfoutput=1
\usepackage{epsfig}
\usepackage{amsmath}

\title{Spectral Sequences and Vacua in $\mathcal N= 2$ Gauged Linear Quantum Mechanics with Potentials}

\author{Kenny Wong\\
Department of Applied Mathematics and Theoretical Physics, \\
Centre for Mathematical Sciences, \\
University of Cambridge, \\
Cambridge, CB3 0WA, UK\\{\tt k.wong@damtp.cam.ac.uk}}

\abstract{We study the behaviour of supersymmetric ground states in a class of one-dimensional $\mathcal N = 2$ abelian gauged linear sigma models, including theories for which the target space is a complete intersection in projective space, and more generally, models with an interaction term introduced by Herbst, Hori and Page in which the vacua correspond to elements of hypercohomology groups of complexes of sheaves. Combining physical insights from recent work by Hori, Kim and Yi with the use of spectral sequences, we propose a way to reconcile the non-linear sigma model description, valid deep within a geometric phase, with the effective Coulomb branch description, valid near a phase boundary. This leads to a physical interpretation of the hypercohomology groups from the perspective of the Coulomb branch, as well as an interpretation for the spectral sequences used to compute them.}

\begin{document}
\maketitle
\flushbottom

\section{Introduction}

\paragraph{}
Gauged linear sigma models \cite{wittenphases} have been at the forefront of research at the interface between theoretical physics and geometry for over twenty years, providing an arena for the cross-fertilisation of ideas between these two disciplines.  This has led to constructions of mirror symmetry \cite{mirror, mirror2}, derivations of derived equivalences \cite{hhp, willwindow, jockers4}, exact computations of Gromov-Witten invariants, \cite{exact1, exact2, exact3, jockers2, jockers3}, insights into subtle aspects of determinantal varieties \cite{nonabelian, jockers1, willpfaffian} and noncommutative resolutions \cite{sharpe1, sharpe2, sharpe3, kuznetsov}, and more recently, a mathematically rigorous formulation of GLSMs  \cite{rigorous}. All of these achievements pertain to two-dimensional theories with four real supercharges. There has been relatively less emphasis on gauged linear models in one dimension, or with two supercharges, and one may well ask if there are any interesting geometric properties that are unique to these latter theories.

\paragraph{} 
A key feature of gauged linear sigma models is that the space of classical vacua depends on various background parameters, such as the Fayet-Iliopoulos parameter $\zeta$. Higgs branches of classical vacua are spanned by chiral multiplets, and correspond to symplectic quotients of $\mathbb C^N$ by the complexified gauge group. The Fayet-Iliopoulos parameter $\zeta$ acts as a stability parameter for the symplectic quotient, and as $\zeta$ crosses certain singular loci in its moduli space, the geometry of these symplectic quotients changes. In two-dimensional abelian gauged linear sigma models, the Fayet-Iliopoulos parameter $\zeta$ pairs with the $\theta$ angle to form a complex parameter, and quantum effects lift the singular locus to a point in the $\zeta + \frac {i}{2\pi} \theta $ moduli space. This means that, by avoiding these singularities in the moduli space, one can pass smoothly from one phase to another. One-dimensional abelian gauged linear models are inherently different: $\zeta$ is a \emph{real} parameter, and the wall between phases cannot be avoided. Therefore, it is possible for the Witten index to jump as $\zeta $ crosses a wall.

\paragraph{}
In this article, we focus on abelian  gauged linear quantum mechanics with positively charged chiral multiplets. In the $\zeta \to \infty $ limit, the supersymmetric vacua in such models localise on  a compact Higgs branch, and the theory is well described by a non-linear sigma model. Typically the number of supersymmetric vacua is given by  the dimensions of certain cohomology groups associated with the target space. In the opposite regime, $ \zeta < 0 $, the Higgs branch is empty, and these supersymmetric vacua disappear \cite{denef, moredenef, hky}. What happens in the intermediate regime, as $ \zeta $ is reduced from infinity to zero, is interesting.  At the phase boundary $\zeta = 0$, a new branch of classical vacua emerges, known as the Coulomb branch. This is a non-compact region, spanned by the vector multiplet scalar. As $\zeta$ approaches zero, the wavefunctions of the supersymmetric vacua on the compact Higgs branch move along the non-compact Coulomb branch, and when $\zeta $ reaches zero, they disappear at infinity, far along the Coulomb branch. This kind of phenomenon, in which normalisable wavefunctions become non-normalisable at infinity when certain parameters cross walls in a parameter space, has also been observed in the moduli space description of BPS solitons (\cite{sw1, sw2,framed,kswall}; see also \cite{sungjay, me1, me2} for a few of the simplest examples of this interpretation).

\paragraph{}
What is special about one-dimensional abelian $\mathcal N = 2$ models is that the Coulomb branch is spanned by a single, real variable, $\sigma \in \mathbb R$. Thus, the Coulomb branch has \emph{two} distinct regions at infinity, $\sigma \to - \infty$ and $\sigma \to + \infty$. It is therefore reasonable to ask the following question: \emph{As we vary the FI parameter from infinity to zero, how many of the Higgs branch vacua disappear at the $\sigma \to - \infty$ end of the Coulomb branch, and how many disappear at the $\sigma \to + \infty$ end? Can we relate these numbers to geometric properties of the Higgs branch?} It is worth stressing that this question is unique to $\mathcal N = 2$ quantum mechanics; it would make no sense to ask this question, say, in $\mathcal N = 4$ quantum mechanics, or two-dimensional $\mathcal N = (2,2)$ models, where the boundary of the Coulomb branch has only one connected component.

\paragraph{}
The aim of this article is to  study the behaviour of supersymmetric vacua on the Coulomb branch for a class of supersymmetric deformations that are unique to $\mathcal N = 2$ models and have clear and natural geometric interpretations. The simplest among the models we study are free theories with a Wilson line; the Coulomb branch analysis for these theories has already been carried out in \cite{hky}. We may deform these free theories by introducing Fermi multiplets with $E$-term potentials, and the theories we obtain in this way reduce in the $\zeta \to \infty$ limit to non-linear sigma models on complete intersections in projective space. These $E$-term deformations are themselves a subclass of a more general class of deformations, associated to complexes of vector bundles on projective space \cite{hhp}.

\paragraph{}
For this general class of deformations, the supersymmetric vacua in the non-linear model describing the $\zeta \to \infty$ limit of the theory are given by the hypercohomology groups of the associated complex of vector bundles. Our goal is to follow the fate of these vacua as they escape to infinity along either of the Coulomb branch directions in the limit $\zeta \to 0 $. Of course, one faces many difficulties. For instance, the presence of non-linear potential terms makes it impossible to compute analytically the effective action on the Coulomb branch, even at lowest order in the derivative expansion. Furthermore, while techniques for computing Witten indices do exist (see e.g. \cite{hky, otherloc1, otherloc2} for a few examples of localisation calculations of indices in one-dimensional theories), these do not quite suffice for our purposes because we wish to characterise the wavefunctions of individual states.

\paragraph{}
Nonetheless, in spite of these difficulties, we shall see that it is possible to put forward a consistent proposal as to how the supersymmetric vacua behave on the Coulomb branch. Rather than performing an index calculation, we study the $Q$-cohomology classes of quantum states using spectral sequences. Spectral sequences are a tool from homological algebra, and one of their main applications is to simplify calculations of homology groups for bigraded complexes. For us, the terms in these complexes represent physical states in the quantum Hilbert space, and the differentials represent supercharges, while the homology groups represent spaces of supersymmetric vacua. In fact, the use of spectral sequences for computing hypercohomology groups for complexes is standard practice, and this corresponds precisely to the counting of supersymmetric vacua in the non-linear model on the Higgs branch. What is a little more surprising is that one can also define a pair of spectral sequences that simplify the analysis of the theory on each of the two regions of the Coulomb branch. The interplay between these three spectral sequences is the key to understanding the relationship between the Higgs and Coulomb branch descriptions of the gauged linear sigma model.

\paragraph{}
This method of analysis is interesting from a mathematical point of view. Normally, when one calculates the hypercohomology of a complex $(\mathcal F^\bullet, f^\bullet)$ using the spectral sequence
\begin{eqnarray}
E_2^{p,q} = H_{f}^p ( H^q (F^\bullet)) \implies E_\infty^{p,q} = \mathbb H^{p+q} (\mathbb F^\bullet) \nonumber
\end{eqnarray}
one is ultimately interested in computing the direct sum, $\oplus_{p+q = r} E_\infty^{p,q}$, as this is what determines the hypercohomology group $\mathbb H^r (\mathcal F^\bullet)$. Our physical analysis, however, suggests that, under certain assumptions, the individual vector spaces $E_\infty^{p,q}$ are significant in their own right, because they tell us how many supersymmetric vacua become non-normalisable at each of the two ends of the Coulomb branch as the Fayet-Iliopoulos parameter $\zeta$ tends to zero. We also find interpretations for $E_2^{p,q}$, and for some of the differentials in the spectral sequence, which relate to properties of the quantum states in the matter sector of the theory.

\paragraph{}
Recently, a class of $\mathcal N = 4$ quiver quantum mechanics was solved numerically \cite{computer}. It would be fascinating to test the proposals in this article about $\mathcal N = 2$ quantum mechanics in the same way.

\paragraph{}
The structure of the paper is as follows. In Sections 2.1 and 2.2, we describe the class of $\mathcal N = 2$ theories that we will study. In Sections 2.3 and 2.4, we explain how spectral sequences can be used to simplify the counting of supersymmetric vacua in the Higgs branch description of the theory deep inside the geometric phase; this material will serve as a basis for comparison later on. The bulk of the analysis is in Section 3, where we study the Coulomb branch description of the theory near the phase boundary, leading up to the main proposal stated in Section 3.5.  Finally, we discuss generalisations and future directions in Section 4.

\section{Supersymmetric Vacua: Higgs Branch Analysis} 

\subsection{Wilson Lines, Potentials and Hypercohomology}

\paragraph{}Throughout this article, we focus on one-dimensional $\mathcal N = 2$ gauged linear models  with a $U(1)$ vector multiplet $(u_t, \sigma, \lambda, D)$ and $N+1$ chiral multiplets $(\phi_i, \psi_i)$, $ i = 0, ... , N$, of charge one. All the theories we study contain the following terms in their Lagrangian:\footnote{The supersymmetry transformations are: $\delta \phi = \psi \epsilon,$ $ \delta \psi = i (D_t \phi + i \sigma \phi) \bar\epsilon$, $ \delta u_t = - \delta \sigma = - \frac i 2 \bar \lambda \epsilon + \frac i 2 \bar \epsilon \lambda,$ $ \delta \lambda = (\dot \sigma + i D) \epsilon, $ $\delta D = - \frac 1 2 \dot {\bar \lambda } \epsilon - \frac 1 2 \bar\epsilon \dot \lambda $
.}
\begin{eqnarray}
L_{\rm GLSM} &= &\frac 1 {2e^2} ( \dot \sigma ^2 + i \bar \lambda \dot \lambda + D^2 ) - \zeta D \nonumber \\
&& + D_t \bar \phi_i D_t \phi_i +  \bar \phi_i (D - \sigma^2 ) \phi_i + i \bar\psi_i D_t \psi_i + \bar \psi_i  \sigma \psi_i - i \bar \phi_i \lambda \psi_i + i \bar\psi_i \bar \lambda \phi_i.  \nonumber
\end{eqnarray}
The covariant derivatives are defined as
\begin{eqnarray}
D_t \phi_i = \dot \phi_i + i u_t \phi_i , \ \ \ \ \ \ \ \ \ \ \  D_t \psi_i = \dot \psi_i + i u_t \psi_i. \nonumber
\end{eqnarray}

\paragraph{}
When $\zeta > 0 $, the classical vacua lie on the Higgs branch,
\begin{eqnarray}
\bar\phi_i \phi_i  = \zeta, \ \ \ \ \  \ \ \ \ \  \sigma = 0 .\nonumber
\end{eqnarray}
The $U(1)$ gauge symmetry is spontaneously broken on the Higgs branch, and the action of this $U(1)$ group identifies
\begin{eqnarray}
\phi_i \sim e^{i \alpha} \phi_i. \nonumber
\end{eqnarray}
Geometrically, the Higgs branch is the complex projective space $\mathbb P^N$. In the $\zeta \to \infty$ limit, the support of the wavefunctions of the quantum ground states localise to the Higgs branch. Since the vector multiplet fields become infinitely massive in this limit, they can be integrated out, giving a non-linear sigma model on the $\mathbb P^N$ target space \cite{wittenphases}. This regime will be the focus of the present section.

\paragraph{}
In the limit $\zeta \to 0$, the Higgs branch shrinks to zero size. But precisely in this $\zeta \to 0$ limit, the Coulomb branch of classical vacua emerges,
\begin{eqnarray}
\phi_i = 0 , \ \ \ \ \ \ \ \ \, \sigma \in \mathbb R. \nonumber
\end{eqnarray}
The Coulomb branch is a real line, attaching to the Higgs branch at the point $\sigma = 0$, and it has two asymptotic directions, $\sigma \to - \infty$ and $\sigma \to + \infty$. For $\zeta$ close to zero, the description of the theory as a non-linear model on the Higgs branch is not appropriate: the theory is better described by an effective theory for the Coulomb branch variable $\sigma$; the analysis of this Coulomb branch description will occupy us in Section 3.

\paragraph{}
One can add a variety of terms to this model, consistent with $\mathcal N = 2$ supersymmetry. The simplest of them is a Wilson line, or equivalently, a Chern-Simon term for the $U(1)$ gauge field,
\begin{eqnarray}
& L_{\rm Wilson} =  - \left(k +  \frac {N+1} 2 \right)(u_t + \sigma) .\nonumber
\end{eqnarray} 
We take $k \in \mathbb Z$ so that the theory does not suffer from the parity anomaly after quantisation. It should be emphasised that this Wilson line term is unique to $\mathcal N = 2$ models in one dimension: it is not invariant under $\mathcal N = 4$ supersymmetry, nor is it invariant under two-dimensional Lorentz symmetry. Since the field $\sigma $ appears in the Wilson line action, the Wilson line gives rise to a linear potential along the Coulomb branch, breaking the $\sigma \mapsto - \sigma$ symmetry. It is worth pointing out an application of this $\mathcal N = 2$ Wilson line deformation: it describes the moduli space dynamics of BPS vortices in three-dimensional Chern-Simons theories \cite{collie}.

\paragraph{}
One can derive the non-linear sigma model description of the $\zeta \to \infty$ limit in the usual way, by eliminating the variables $u_t$ and $\sigma$ using their equations of motion. In doing so, the Wilson line term provides an additional contribution to the non-linear sigma model action: a term with first-order time derivatives in the bosons (see for instance \cite{hhp, hky}). This is a connection on the line bundle $\mathcal O_{\mathbb P^N} (k + \frac {N+1} 2)$. The Hilbert space for this non-linear model consists of $(0,q)$-form valued sections of the bundle $\mathcal  O_{\mathbb P^N} (k + \frac {N+1} 2) \otimes K_{\mathbb P^N}^{1/2}  \cong \mathcal O_{\mathbb P^N}(k)$, and the supercharge is the Dolbeault operator $\bar \partial_{\mathcal O_{\mathbb P^N}(k)}$. The result is that the supersymmetric vacua of $R$-charge $r$ are in correspondence with elements of the sheaf cohomology group,
\begin{eqnarray}
& \mathcal H^r_0 \cong H^r (\mathbb P^N,  \mathcal  O_{\mathbb P^N} (k )) . \nonumber
\end{eqnarray}
(We adopt the convention that operators in the Lagrangian are ordered symmetrically; the factor of $K_{\mathbb P^N}^{1/2} \cong \mathcal O_{\mathbb P^N} (- \frac {N+1} 2 )$ can then be traced to normal ordering \cite{hky}.)

\paragraph{}
Another way in which we can deform this theory while preserving $\mathcal N = 2$ supersymmetry is to introduce Fermi multiplets with $E$-term potentials. This is a natural deformation, and lends itself to a simple interpretation in terms of complete intersections in projective space. Although we could describe this deformation now, it will prove more helpful for our purposes to first describe a more general, but possibly less familiar, class  of $\mathcal N = 2$ deformations, first introduced in \cite{hhp} as boundary conditions for two-dimensional $\mathcal N = (2,2)$ theories. (We shall revisit Fermi multiplets in the language of \cite{hhp} in Section 2.2.)

\paragraph{}
To construct the $\mathcal N = 2$ deformation of \cite{hhp}, we introduce a Chan-Paton space\footnote{The terminology is inherited from two-dimensional theories with boundaries. Each basis vector for the Chan-Paton space represents a distinct Fock space vacuum, and correspondingly the Lagrangian and Hamiltonian are matrix-valued quantities.}
\begin{eqnarray}
\mathcal V_{\rm CP} = \underset{p \in \mathbb Z} \oplus \mathcal V^p . \nonumber
\end{eqnarray}
We then replace our Wilson line with a matrix-valued Lagrangian,
\begin{eqnarray}
& L_{\rm Wilson} = - \left(\rho + {\rm diag}( \frac {N+1} 2 , ... , \frac {N+1} 2)\right)  (u_t + \sigma). \nonumber
\end{eqnarray}
The matrix $\rho$ takes the block diagonal form,
\begin{eqnarray}
\rho = \left( \begin{array}{c|c|c|c|c} & \vdots & \vdots & \vdots & \\ \hline \dots & \rho^{p-1} & 0 & 0 & \dots \\ \hline  \dots & 0 &\  \  \rho^p \  \ & 0 & \dots \\ \hline   \dots & 0 & 0 & \rho^{p+1}  \\ \hline & \vdots  &\vdots & \vdots &   \end{array} \right),
\nonumber
\end{eqnarray}
where $\rho^p$, the block acting on the $D^p$-dimensional Chan-Paton space $\mathcal V^p$ is of the form
\begin{eqnarray}
\rho^p = \left(\begin{array}{ccc} k_1^p  & 0 & 0 \\ 0 & \  \ \ddots \  \ & 0  \\0  & 0 & k_{D^p}^p  \nonumber \end{array} \right),
\end{eqnarray}
with $k_j^p \in \mathbb Z$.
One can also add a potential term,
\begin{eqnarray}
L_{\rm Pot} = - \{ f, f^\dagger \} +  \psi_i \partial_i f - \bar\psi_i \partial_{\bar i} f^\dagger, \nonumber
\end{eqnarray}
where $f$ is a matrix of block diagonal form,
\begin{eqnarray}
f(\phi) =  \left( \begin{array}{c|c|c|c|c} & \vdots & \vdots & \vdots & \\ \hline \  \dots \  & 0 & 0 & 0 & \dots \\ \hline \dots &  f^{p-1}(\phi)    & 0 & 0 &   \dots \\ \hline   \dots & 0 & \  f^{p}(\phi) \   & \ \ \  \ 0 \  \  \ \  & \ \dots \ \\ \hline & \vdots  &\vdots & \vdots &   \end{array} \right).
\nonumber
\end{eqnarray}
Each $f^p(\phi) $ is a $D^{p+1} \times D^p$ matrix,
\begin{eqnarray}
f^p(\phi)  = \left(\begin{array}{ccc} f^p_{11}(\phi) & \dots  & f^p_{1 D^p}(\phi) \\ \vdots & \ \ \ \ \ \ddots \  \ \ \ \ &  \vdots \\ f^p_{D^{p+1} 1}(\phi) & \dots  & f^p_{D^{p+1} D^p}  (\phi)  \nonumber \end{array} \right)
\end{eqnarray}
such that the entry $f^p_{jj'}$ is a homogeneous polynomial in $\phi_0, ... , \phi_N$ of degree $k_j^{p+1} - k_{j'}^p$. The matrices $f^p$ are required to obey
\begin{eqnarray}
f^{p+1} \circ f^p = 0 \nonumber.
\end{eqnarray}
Since the endomorphism $f$ maps even Chan-Paton states to odd Chan-Paton states and vice versa, it anticommutes with the fermionic fields in the theory.

\paragraph{}
As explained in \cite{hhp, hky}, the Hilbert space for the non-linear model is the direct sum of the vector spaces of smooth $(0,q)$-forms on the bundles $\mathcal O(k_j^p)$,
\begin{eqnarray}
\mathcal H = \oplus_{p,q} \mathcal H^{p,q}, \ \ \ \ \ \ \ \ \ \ \  \mathcal H^{p,q} = \Gamma \left[  \mathcal A^{0, q} \left( \oplus_{j=1}^{D^p} \mathcal O_{\mathbb P^N} (k_j^p) \right) \right] . \nonumber
\end{eqnarray}
(The notation $\mathcal A^{0, q} (E)$ denotes the sheaf of smooth $(0,q)$-form-valued sections of the vector bunde $E$, and $\Gamma$ denotes taking global sections.) The supercharge operator acts on $\mathcal H^{p,q}$ as
\begin{eqnarray}
Q^{p,q} = \bar \partial + (-1)^q f^p. \nonumber
\end{eqnarray}
One can view $\mathcal H^{p,q}$ as a bigraded complex of vector spaces. The two terms that make up the supercharge operator act as differentials for this complex, with
\begin{eqnarray}
\bar \partial : \mathcal H^{p,q} \to \mathcal H^{p,q+1}, \ \ \ \ \ \ \   f^p : \mathcal H^{p,q} \to \mathcal H^{p+1,q}. \nonumber
\end{eqnarray}
The space of states of $R$-charge $r$ form the vector space,
\begin{eqnarray}
\mathcal H^r = \underset{p+q = r} {\oplus} \mathcal H^{p,q}. \nonumber 
\end{eqnarray}
These spaces $\mathcal H^r$ form a complex with a single grading, and the differential for this complex is
\begin{eqnarray}
Q^r : \mathcal H^r \to \mathcal H^{r+1}, \nonumber
\end{eqnarray}
where
\begin{eqnarray}
Q^r = \underset{p+q=r}{\oplus} Q^{p,q}. \nonumber
\end{eqnarray}
Finally, the supersymmetric ground states of $R$-charge $r$ are given by the cohomology of this single-graded complex,
\begin{eqnarray}
\mathcal H^r_0 \cong H_Q^r (\mathcal H^\bullet ). \nonumber
\end{eqnarray}
This is nothing other than the hypercohomology group\footnote{An algebraically-minded reader may be more accustomed to seeing hypercohomology defined by taking injective resolutions, or Cech resolutions on an open affine cover. Here, we work in the analytic category, viewing our target space $\mathbb P^N$ as a complex manifold. Our definition of hypercohomology, given in terms of resolutions by the acyclic sheaves $\mathcal A^{0,q}\left(\oplus_j O_{\mathbb P^N} (k_j^p)\right)$, is equivalent to the algebraic definition.} $\mathbb H^r (\mathcal F^\bullet)$ for the complex of vector bundles,
\begin{eqnarray}
\mathcal F^\bullet = \dots \to \oplus_{j=1}^{D^{p-1}} \mathcal O_{\mathbb P^N} (k_j^{p-1}) \overset{f^{p-1}} {\rightarrow} \oplus_{j=1}^{D^{p}} \mathcal O_{\mathbb P^N} (k_j^{p}) \overset{f^{p}} {\rightarrow} \oplus_{j=1}^{D^{p+1}} \mathcal O_{\mathbb P^N} (k_j^{p+1})  \to \dots \nonumber
\end{eqnarray}

\paragraph{}
Readers familiar with D-brane categories will recognise the hypercohomology group $\mathbb H^r (\mathcal F^\bullet)$  as the group of morphisms from the trivial line bundle $\mathcal O_{\mathbb P^N}$ to the complex $\mathcal F^\bullet$ in the derived category of $\mathbb P^N$.
\begin{eqnarray}
\mathbb H^r (\mathcal F^\bullet ) \cong  {\rm Hom}_{\mathbb D^b (\mathbb P^N)} ^r (\mathcal O_{\mathbb P^N}, \mathcal F^\bullet ). \nonumber
\end{eqnarray}
The derived category $\mathbb D^b (\mathbb P^N)$ is the category whose objects are bounded complexes of coherent sheaves on $\mathbb P^N$ and whose morphisms are chain maps, and quasi-isomorphisms (i.e. chain maps inducing isomorphisms on homology) are invertible. Derived categories appear on the B-model side of homological mirror symmetry \cite{kontsevich}, and provide a description for B-branes in string theory (see \cite{aspinwall} for a review and references). This correspondence between categories and branes has led to fruitful interactions between mathematics and physics, a notable example being the notion of stability conditions for derived categories \cite{sharpe, douglas, bridgeland1, bridgeland2}. (See also \cite{atwist1, atwist2} for a connection between hypercohomology and A-twisted Landau-Ginzburg theories.)

\paragraph{}
Of course, not every coherent sheaf is the direct sum of line bundles, so some objects in $\mathbb D^b (\mathbb P^N)$ are not directly associated to an $\mathcal N = 2$ quantum mechanics of the form described. However, all complexes of coherent sheaves on $\mathbb P^N$ are quasi-isomorphic to complexes of direct sums of line bundles, since $\langle \mathcal O_{\mathbb P^N}, ... , \mathcal O_{\mathbb P^N} (N) \rangle$ is an exceptional collection for $\mathbb D^b (\mathbb P^N)$ \cite{beilinson1, beilinson2}. We will exploit this fact to construct theories with interesting target spaces in Section 2.2, and also in the examples in Section 2.4.

\subsection{Fermi Multiplets and Complete Intersections in Projective Space}

\paragraph{}
As promised, we now explain how $\mathcal N = 2$ Fermi multiplets form a subclass of the interactions introduced by \cite{hhp} and reviewed in Section 2.1. For simplicity, let us consider what happens when we add a single Fermi multiplet $\eta$ of charge $d$ to our basic Lagrangian $L_{\rm GLSM}$. The new terms are\footnote{The supersymmetry transformation of $\eta$ is $\delta \eta = E \bar\epsilon$.}
\begin{eqnarray}
 L_{\rm Fermi} = i \bar \eta \dot \eta - d(u_t + \sigma) \bar \eta \eta -  | E|^2 - \partial_i E \bar \eta \psi_i -  \partial_{\bar i} \bar E \bar \psi_i \eta, \nonumber
\end{eqnarray}
where the superpotential $E(\phi_0, ..., \phi_N)$ is a homogeneous polynomial of degree $d$. We also add an anomaly-free Wilson line,
\begin{eqnarray}
& L_{\rm Wilson}  = - \left(k + \frac {N+1-d} 2 \right) (u_t + \sigma ) . \ \ \ \ \ \ \  \nonumber
\end{eqnarray}

\paragraph{}
Let us assume that the gradient vector $\partial_i E$ is non-vanishing on the hypersurface $X = (E(x_0, \dots , x_N)= 0) \subset \mathbb P^N$. Then $X$ is smooth, and in the limit $\zeta \to \infty$, the theory is described by an $\mathcal N = 2$ non-linear sigma model on $X$. Indeed, the $| E |^2$ potential term constrains the bosonic degrees of freedom within this hypersurface, and the $\partial_i E \bar \eta \psi_i$ Yukawa term ensures that the massless fermionic excitations are the $\psi_i$ modes orthogonal to $\partial_i E$, that is, the $\psi_i$ modes tangent to the hypersurface. One then expects that the space of supersymmetric vacua of $R$-charge $r$ is
\begin{eqnarray}
\mathcal H^r_0 \cong H^r ( X,  \mathcal O_X (k +& \frac {N+1-d} 2 ) \otimes K_X^{1/2} )  = H^r (X, O_X(k)), \nonumber
\end{eqnarray}
since $K_X = \mathcal O_{\mathbb P^N} (d) \otimes K_{\mathbb P^N} |_X =  \mathcal O_X (d - N-1)$ by the adjunction formula.

\paragraph{}
To relate this model to the class of models discussed in Section 2.1, we note that the kinetic term for $\eta$ enforces the canonical anticommutation relation,
\begin{eqnarray}
\{ \eta , \bar \eta \} = 1. \nonumber
\end{eqnarray}
As noted in \cite{hhp, hky}, integrating out $\eta$ generates a Chan-Paton space,
\begin{eqnarray}
\mathcal V = \mathcal V^{-1} \oplus \mathcal V^0 = \mathbb C | 0 \rangle \oplus \mathbb C \bar\eta | 0 \rangle.  \nonumber
\end{eqnarray}
After adding the normal ordering constant $-\frac 1 2 $ to $\bar \eta \eta$, it is easy to see that the remaining terms in $ L_{\rm Fermi} +  L_{\rm Wilson}$ act on this Chan-Paton space in a way that is equivalent to the Lagrangian below:
\begin{eqnarray}
& L_{\rm Wilson + Pot } = - \left(\rho + {\rm diag}( \frac {N+1} 2 , ... , \frac {N+1} 2)\right)  (u_t + \sigma)  - \{ f, f^\dagger \} +  \psi_i \partial_i f -  \bar\psi_i \partial_{\bar i} f^\dagger. \nonumber
\end{eqnarray}
Here, the matrices $\rho$ and $f$ are
\begin{eqnarray}
\rho = \left( \begin{array}{cc} k-d & \  \ 0  \\ 0 & \ \  k \end{array} \right), \ \ \ \ \ \  f = \left( \begin{array}{cc} 0 &  \ 0 \\  E(\phi) &  \ 0  \end{array} \right) . \nonumber 
\end{eqnarray}
This data corresponds to the two-term complex,
\begin{eqnarray}
\mathcal F^\bullet =  \mathcal O_{\mathbb P^N} (k-d) \overset{E}{\xrightarrow{\hspace*{0.5cm}}} \underline{\mathcal O_{\mathbb P^N}(k ) }. \nonumber
\end{eqnarray}
Therefore, the space of supersymmetric ground states of $R$-charge $r$, in the non-linear model limit $\zeta \to \infty$, is
\begin{eqnarray}
\mathcal H^r_0 = \mathbb H^r \left( O_{\mathbb P^N} (k-d) \overset{E}{\to} \underline{\mathcal O_{\mathbb P^N}(k) }\right). \nonumber 
\end{eqnarray}
But in view of the short exact sequence\footnote{If we relax the condition that $\partial_i E$ is non-vanishing, then $X$ may be a singular hypersurface, but the short exact sequence still holds. For instance, if $E(\phi_0, \phi_1)  = \phi_0^2$, then $X$ is the non-reduced scheme ${\rm Proj}\  \mathbb C [x_0, x_1] / (x_0^2)$ and $i_\star \mathcal O_X$ is a skyscraper sheaf of degree two  supported at $ [0:1] \in \mathbb P^1$. Similarly, for complete intersections, the condition that $\partial_i E_j$ is of maximal rank can be relaxed to the condition that $E_1, \dots E_M$ form a regular sequence, i.e. each $E_j$ is a non-zero-divisor in $\mathbb C[x_0, \dots, x_N] / (E_1, \dots E_{j-1})$.},
\begin{eqnarray}
0 \to  O_{\mathbb P^N} (k-d) \overset{E}{\to} \mathcal O_{\mathbb P^N}(k)  \to i_\star \mathcal O_X (k) \to 0 , \nonumber
\end{eqnarray}
where $i: X \hookrightarrow \mathbb P^N$ is the closed immersion of $X$ into $\mathbb P^N$, it is clear that the complex $\mathcal F^\bullet$ is quasi-isomorphic to $i_\star\mathcal O_X (k)$, so the space of supersymmetric ground states is isomorphic to
\begin{eqnarray}
\mathcal H^r_0 = H^r (\mathbb P^N, i_\star \mathcal O_X(k)) = H^r (X, \mathcal O_X (k)), \nonumber
\end{eqnarray}
which is what we expect.

\paragraph{}
This construction generalises straightforwardly for complete intersections in projective space. One introduces a collection of Fermi multiplets $\eta_j$ of charge $d_j$, with $j = 1, ... , M$, and adds potentials $E_j(\phi_0, ..., \phi_N)$ given by homogeneous degree $d_j$ polynomials such that the Hessian matrix $\partial_i E_j$ is of maximal rank on $X = (E_1 = ... = E_M   = 0) \subset \mathbb P^N$. One also introduces a Wilson line of charge $k + \frac 1 2 (N+1 -  \sum_{j=1}^M d_j)$.  The target space of the non-linear sigma model is the smooth complete intersection $X$. Integrating out the Fermi multiplets generates a $2^M$-dimensional Chan-Paton space, and the associated complex of vector bundles is a locally free resolution of the sheaf $i_\star O_X(k)$, where again, $i : X \hookrightarrow \mathbb P^N$ is the natural embedding.

\subsection{Spectral Sequences}

\paragraph{}
Spectral sequences are the main analytical tool in this article. We will encounter three spectral sequences altogether: one for the Higgs branch, and one for each of the two regions of the Coulomb branch. The Higgs branch spectral sequence is nothing other than the standard spectral sequence used to simplify the computation of  hypercohomology groups. Since our subsequent discussion relies on detailed comparisons between this hypercohomology spectral sequence and the two Coulomb branch spectral sequences, we will now take some time to explain how the hypercohomology spectral sequence works, highlighting the features that will become relevant later on.

\paragraph{}
The general situation (in both the Higgs and the Coulomb branch pictures) is as follows: we have a bigraded complex of abelian groups $K^{p,q}$ with differentials
\begin{eqnarray}
\delta^{p,q} : K^{p,q} \to K^{p, q+1}, \ \ \ \ \ \ \ \ \ \ \ \ {\delta'}^{p,q} : K^{p,q} \to K^{p+1, q}, \nonumber
\end{eqnarray}
satisfying
\begin{eqnarray}
&& \delta^{p,q+1} \circ \delta^{p,q} = 0, \ \ \ \ \ \ \ \ {\delta'}^{p+1, q} \circ {\delta'}^{p,q} = 0, \nonumber \\  && \ \ \ \ \ \   \delta^{p+1 . q} \circ {\delta'}^{p,q} + {\delta'}^{p,q+1} \circ \delta^{p,q} = 0 . \nonumber
\end{eqnarray}
From the bigraded complex $K^{p,q}$, we can form a complex with a single grading $r$, whose terms are
\begin{eqnarray}
K^r = \underset{p+q = r} \oplus K^{p,q} \nonumber
\end{eqnarray}
and whose differential is
\begin{eqnarray}
D^r = \underset{p+q = r} {\oplus} \left( \delta^{p,q} + {\delta'}^{p,q} \right). \nonumber
\end{eqnarray}
In the context of our physical model, these abelian groups will always represent quantum states. The $r$ grading  will always denote the $R$-charge in some sector of the theory, and the differential $D^r$ will represent the action of one of the supercharges.

\paragraph{}
Associated to the double complex $K^{p,q}$ is a spectral sequence, that is, a sequence of abelian groups $E_s^{p,q}$ for $s \geq 0$, and a sequence of group homomorphisms
\begin{eqnarray}
d_s^{p,q} : E_s^{p,q} \to E_s^{p+s, q-s+1}, \nonumber
\end{eqnarray}
which square to zero,
\begin{eqnarray}
d_s^{p+s, q-s+1} \circ d_s^{p,q} = 0,
\nonumber
\end{eqnarray}
such that each successive term in the sequence is the $d_s$-homology of the previous term,
\begin{eqnarray}
E_{s+1}^{p,q} = \frac{{\rm Ker } \left(d_s^{p,q} : E_s^{p,q} \to E_s^{p+s, q-s+1} \right) }{{\rm Im }\left( d_{s}^{p-s,q+s-1} : E_s^{p-s, q+s-1} \to E_s^{p.q} \right) }.  \nonumber
\end{eqnarray}

\paragraph{}
The first few terms in this spectral sequence can be given in simple terms. The $s=0$ term is the original bigraded complex,
\begin{eqnarray}
E_0^{p,q} = K^{p,q}, \nonumber
\end{eqnarray} 
and the $s=0$ differential is the vertical differential,
\begin{eqnarray}
d_0^{p,q} = \delta^{p,q} .\nonumber
\end{eqnarray}
Taking homology in $d_0$, one finds that the $s = 1$ term of the spectral sequence is
\begin{eqnarray}
E_1^{p,q} = H_{\delta}^q ( K^{p,\bullet}). \nonumber
\end{eqnarray}
Next, for any element $[a] \in E_1^{p,q}$ represented by an element $a \in K^{p,q}$, we define the action of the $d_1^{p,q}$ differential by
\begin{eqnarray}
d^{p,q}_1 \left( [a] \right) = [{\delta'}^{p,q} (a)] \in E_1^{p+1, q}. \nonumber
\end{eqnarray}
(One can check from the properties of $\delta$ and $\delta ' $ that this differential is well-defined.) Taking homology now in $d_1^{p,q}$, one finds that the $s = 2$ term of the spectral sequence is
\begin{eqnarray}
E_2^{p,q} = H_{\delta '}^p \left( H_{\delta}^q (K^{\bullet, \bullet}) \right). \nonumber
\end{eqnarray}
What is useful about the spectral sequence is that it eventually terminates, that is, for sufficiently large $s$,
\begin{eqnarray}
E_s^{p,q} = E_{s+1}^{p,q} = E_{s+2}^{p,q} = \dots : = E_\infty^{p,q}, \ \ \  \ \  \ \ \ \ \ \ \  \ \ \  s \gg 0 . \nonumber
\end{eqnarray}
Moreover, $E_\infty^{p,q}$ gives the homology of the  single-graded complex $(K^\bullet, D^\bullet)$,
\begin{eqnarray}
H_D^r (K^\bullet ) \cong \underset{p+q  = r}{ \oplus} E_\infty^{p,q}, \nonumber
\end{eqnarray}
which, in our physical setup, is precisely the space of supersymmetric vacua that we wish to compute. (In certain other physical contexts, the same kind of spectral sequence computes BPS states \cite{gukov}, BRST cohomology \cite{brst1,brst2,brst3} or massless string states \cite{string}.)

\paragraph{}
Let us apply this formalism to the $\zeta \to \infty$ limit of our theory, in which it reduces to a non-linear model on the Higgs branch. To find the space of supersymmetric vacua, we need to find the hypercohomology of the sheaf complex,
\begin{eqnarray}
\mathcal F^\bullet = \dots \to \oplus_{j=1}^{D^{p-1}} \mathcal O_{\mathbb P^N} (k_j^{p-1}) \overset{f^{p-1}} {\rightarrow} \oplus_{j=1}^{D^{p}} \mathcal O_{\mathbb P^N} (k_j^{p}) \overset{f^{p}} {\rightarrow} \oplus_{j=1}^{D^{p+1}} \mathcal O_{\mathbb P^N} (k_j^{p+1})  \to \dots \nonumber
\end{eqnarray}
This means that the relevant bigraded complex is
\begin{eqnarray}
K^{p,q} = \Gamma \left[ \mathcal A^{0, q} \left( \oplus_{j=1}^{D^p} \mathcal O_{\mathbb P^N} (k_j^p) \right) \right] , \nonumber
\end{eqnarray}
and the differentials are
\begin{eqnarray}
\delta^{p,q} = \bar \partial, \ \ \ \ \ \ \ \ \ {\delta'}^{p,q} = (-1)^q f^p . \nonumber
\end{eqnarray}

\paragraph{}
The $s= 0$ term in the spectral sequence is simply
\begin{eqnarray}
E_0^{p,q} ({\rm NL})= K^{p,q}, \ \ \ \ \  \ \ \ \ \ \ \   d_0^{p,q} = \bar \partial. \nonumber
\end{eqnarray}
We append the abbreviation ${\rm NL}$ to remind the reader that this is the spectral sequence associated with the non-linear model, so as to avoid confusion with two spectral sequences that we will use to study the Coulomb branch picture in Section 3. The $s=1$ term in the spectral sequence is
\begin{eqnarray}
E_1^{p,q} ({\rm NL}) = \oplus_{j=1}^{D_p} H^{q} \left(\mathcal O_{\mathbb P^N} (k_j^p) \right), \ \ \ \  \ \ \ 
d_1^{p,q} = (-1)^q f^{p}, \nonumber
\end{eqnarray}
where $f^{p}$ here denotes the group homomorphism on the cohomology groups $ \oplus_j H^{q} \left(\mathcal O_{\mathbb P^N} (k_j^p) \right)$ induced by the sheaf morphism $f^p$.

\paragraph{}
It is well-known that $H^0(\mathcal O_{\mathbb P^N} (k))$ is isomorphic to the vector space spanned by degree $k$ monomials in $x_0, ... , x_N$. So the $q=0$ row of the $s=1$ term in the spectral sequence is
\begin{eqnarray}
& E_1^{p,0} ({\rm NL})= \oplus_{j=1}^{D_p} \left\langle x_0^{n_0} ... \ x_N^{n_N}  :  \sum_{i=0}^N n_i = k_j^p \right\rangle   , \nonumber
\end{eqnarray}
and the induced map of $f^p$ on cohomology is multiplication by the matrix $f^p_{jj'} (x_0, ... , x_N)$. Taking homology in $f^p$, one obtains the $q = 0$ row of the $s=2$ term in the spectral sequence,
\begin{eqnarray}
& E_2^{p,0} ({\rm NL}) = H^p_{f^p (x_i)} \left(  \oplus_j \langle x_0^{n_0} ... \ x_N^{n_N}  :   \sum_{i=0}^N n_i = k_j^\bullet \rangle \right).  \nonumber
\end{eqnarray}
By Serre duality, $H^N (\mathcal O_{\mathbb P^N} (k)) \cong H^0 (\mathcal O_{\mathbb P^N} (-k - N - 1))^\vee$, so $H^N (\mathcal O_{\mathbb P^N} (k))$  is the dual of the vector space spanned by degree $- k - N - 1$ monomials in $x_0, ... , x_N$. Therefore, the $q = N$ row of the $s = 1$ term in the spectral sequence is
\begin{eqnarray}
& (E_1^{p,N} ({\rm NL}))^\vee =  \oplus_{j=1}^{D_p} \left\langle x_0^{n_0} ... \ x_N^{n_N}  :    \sum_{i=0}^N n_i = -  k_j^p - N - 1 \right\rangle   . \nonumber
\end{eqnarray}
The dual of the differential $d_1^{p,N}$ is multiplication by the hermitian transpose of the matrix $f^p_{jj'} (x_0, ... , x_N)$. Taking homology in this differential, one finds that the $q=N$ row of the $s = 2$ term, after dualising, is
\begin{eqnarray}
& (E_2^{p,N} ({\rm NL}) )^\vee = H^p_{(f^p)^\dagger (x_i)} \left(  \oplus_j \langle x_0^{n_0} ... \ x_N^{n_N}  :  \sum_{i=0}^N n_i = -  k_j^\bullet - N - 1 \rangle \right) . \nonumber
\end{eqnarray}
All other cohomology groups of $O_{\mathbb P^N} (k)$ vanish, so
\begin{eqnarray}
E_1^{p,q} ({\rm NL})= E_2^{p,q} ({\rm NL}) = 0, \ \ \ \ \ \ \ \ \  1 \leq q \leq N-1 . \nonumber
\end{eqnarray}

\paragraph{}
We now make an important observation: since $E_2^{p,q}({\rm NL}) $  is trivial in the rows $1 \leq q \leq N-1$, the differentials $d_2^{p,q}, ... , d_{N}^{p,q}$ must be zero, and so,
\begin{eqnarray}
E_2^{p,q} ({\rm NL}) = ... = E_{N+1}^{p,q} ({\rm NL})  \nonumber
\end{eqnarray}
for all $p$ and $q$. However, it is possible to have a non-trivial differential
\begin{eqnarray}
d_{N+1}^{p,N} : E_{N+1}^{p,N} ({\rm NL}) \to E_{N+1}^{p+N+1, 0}({\rm NL}) . \nonumber
\end{eqnarray}
All subsequent differentials, $d_{N+2}^{p,q}, d_{N+3}^{p,q}, ...$ are again trivial, so the spectral sequence terminates at $s = N+2$.
Thus, we have
\begin{eqnarray}
E_\infty^{p,q} ({\rm NL}) = E_{N+2}^{p,q}({\rm NL}) =
\begin{cases}
E_{N+1}^{p,0} \ / \ {\rm Im } \ d_{N+1}^{p-N-1,N} ({\rm NL}) & q=0; \\
{\rm Ker} \ d_{N+1}^{p,N} ({\rm NL}) & q = N ;\\
0 &  q \notin \{0, N \}.
\end{cases} \nonumber
\end{eqnarray}
The number of supersymmetric vacua of $R$-charge $r$ is
\begin{eqnarray}
{\rm dim } \ \mathcal H^r_0 ({\rm NL}) & = & {\rm dim } \  E_\infty^{r,0} ({\rm NL}) + {\rm dim } \  E_\infty^{r-N,N} ({\rm NL})  \nonumber \\
& = & {\rm dim } \  E_2^{r,0} ({\rm NL}) + {\rm dim } \  E_2^{r-N,N} ({\rm NL}) - {\rm rank } \ d_{N+1}^{r-N-1,N} -  {\rm rank } \  d_{N+1}^{r-N,N}. \nonumber
\end{eqnarray}

\subsection{Examples}

\paragraph{}
We illustrate the main points discussed so far with a few examples. We shall return to these examples when we come to study the Coulomb branch description of the theory near the phase boundary in Section 3; the final example will turn out to be especially important.

\paragraph{(i)}
Let us consider a theory with three chiral multiplets and an interaction term corresponding to the following two-term complex of sheaves on $\mathbb P^2$.
\begin{eqnarray}
\mathcal F^\bullet = \mathcal O_{\mathbb P^2}^{\oplus 2} \overset{{\left( \begin{array}{cc}x_0 & x_1 \\ x_1 & x_2  \end{array}\right) }}{\xrightarrow{\hspace*{1.4cm}}}  \underline{\mathcal O_{\mathbb P^2}(1)}^{\oplus 2}. \nonumber
\end{eqnarray}
The $E_1^{p,q}({\rm NL})$ term of the spectral sequence is obtained by taking cohomologies of the individual sheaves:
\paragraph{}
\begin{center}
\begin{picture}(212,72)
\put (20,0) {$\langle 1 \rangle ^{\oplus 2}$}
\put (80,3){\vector(1,0){50} }
\put (150,0) {$\underline{\langle x_0, x_1, x_2 \rangle}^{\oplus 2}$}
\put (83,18) {$\left( \begin{array}{cc}x_0 & x_1 \\ x_1 & x_2 \end{array} \right)$}

\put (28,50) {0}
\put (80,53){\vector(1,0){50} }
\put (175,50) {0}

\put (28,80) {0}
\put (80,83){\vector(1,0){50} }
\put (175,80) {0}
\end{picture}
\end{center}
We then take homology in the horizontal direction to obtain the $E_2^{p,q}({\rm NL})$ term.
\begin{center}
\begin{picture}(210,60)
\put (68,0) {$0$}
\put (133,0) {$\underline{\mathbb C}^4$}

\put (68,25) {0}
\put (135, 25) {0}

\put (68,50) {0}
\put (135,50) {0}
\end{picture}
\end{center}
Since there is only one non-trivial vector space remaining, there can be no more non-trivial differentials in the spectral sequence, so the spectral sequence terminates at $s = 2$. The number of supersymmetric vacua is given by
\begin{eqnarray}
\mathcal H^r_0({\rm NL}) = \begin{cases} \mathbb C^4 & r = 0 ; \\0 & r \neq 0 . \end{cases} \nonumber
\end{eqnarray}
This result has a nice physical interpretation. Observe that there is a short exact sequence,
\begin{eqnarray}
0 \to  \mathcal O_{\mathbb P^2} ^{\oplus 2} \to   \mathcal O_{\mathbb P^2}(1)^{\oplus 2} \to i_\star \mathcal O_{\mathbb P^1} (3) \to 0,
\nonumber
\end{eqnarray}
where $i : \mathbb P^1 \hookrightarrow \mathbb P^2$ is the second Veronese embedding, $[u_0 : u_1] \mapsto [u_0^2 : u_0u_1 : u_1^2]$. So the complex $\mathcal F^\bullet$ is a resolution for the sheaf $i_\star \mathcal O_{\mathbb P^1} (3)$, and hence $\mathcal F^\bullet$ is quasi-isomorphic to $i_\star \mathcal O_{\mathbb P^1} (3)$ in the derived category of $\mathbb P^2$. So the $\zeta \to \infty$ limit of this theory is an $\mathcal N = 2$ non-linear model for a supersymmetric particle moving on the quadric $i(\mathbb P^1) = (x_0x_2 - x_1^2 = 0) \subset \mathbb P^2$ coupled to a magnetic field with three units of Dirac monopole flux through  $i(\mathbb P^1)$. The supersymmetric vacua can be viewed as lowest Landau level modes on this quadric. The number of supersymmetric vacua agrees with the cohomology of the line bundle $\mathcal O_{\mathbb P^1}(3)$ on $\mathbb P^1$,
\begin{eqnarray}
H^r(\mathbb P^1, \mathcal O_{\mathbb P^1} (3)) = \begin{cases} \mathbb C^4 & r = 0 ; \\ 0 & r = 1 . \end{cases} \nonumber
\end{eqnarray}

\paragraph{(ii)}
Next, we consider another two term complex on $\mathbb P^2$,
\begin{eqnarray}
\mathcal F^\bullet = \mathcal O_{\mathbb P^2}(-5) \overset{{\left( \begin{array}{c} x_0 \\ x_1 \\ x_2  \end{array}\right) }}{\xrightarrow{\hspace*{1.2cm}}}  \underline{\mathcal O_{\mathbb P^2}(-4)}^{\oplus 3}.\nonumber
\end{eqnarray}
In this case, the only non-trivial cohomology groups are the top-degree groups, and it is more convenient to write down the dual vector spaces $\left(E_1^{p,q}({\rm NL})\right)^\vee$ and dual differentials $(d_1^{p,q})^\vee$ after applying Serre duality. These are shown below.
\begin{center}
\begin{picture}(212,87)
\put (28,5) {0}
\put (137,8){\vector(-1,0){63} }
\put (175,5) {\underline{0}}

\put (28,35) {0}
\put (137,38){\vector(-1,0){63} }
\put (175,35) {0}

\put (-72,65) {$\langle x_0^2, x_1^2, x_2^2, x_0x_1, x_0x_2, x_1x_2 \rangle$}
\put (137,68){\vector(-1,0){63} }
\put (153,65) {$\langle x_0, x_1, x_2 \rangle ^{\oplus 3}$}
\put (80,76) {$( \ x_0 \ x_1 \ x_2 \ )$}
\end{picture}
\end{center}
As usual, one takes homology in the horizontal direction to obtain $\left(E_2^{p,q}({\rm NL}) \right)^\vee$, where the spectral sequence terminates:
\begin{center}
\begin{picture}(210,58)
\put (68,0) {$0$}
\put (133,0) {$\underline{0}$}

\put (68,25) {0}
\put (135,25) {0}

\put (68,50) {0}
\put (134,50) {$\mathbb C^3$}
\end{picture}
\end{center}
The supersymmetric ground states are given by
\begin{eqnarray}
\mathcal H^r_0({\rm NL}) = \begin{cases}  \mathbb C^3 & r = 2;  \\ 0 & r \neq 2. \end{cases} \nonumber
\end{eqnarray}
In view of the Euler sequence (tensored by $\mathcal O_{\mathbb P^2}(-5)$),
\begin{eqnarray}
0 \to  \mathcal O_{\mathbb P^2}(-5) \to   \mathcal O_{\mathbb P^2}(-4)^{\oplus 3} \to  \mathcal T_{\mathbb P^2} (-5) \to 0,
\nonumber
\end{eqnarray}
one learns that our complex $\mathcal F^\bullet$ is a resolution for $\mathcal T_{\mathbb P^2}(-5) \cong \mathcal T_{\mathbb P^2} \otimes \mathcal O_{\mathbb P^2} (-5)$. The number of supersymmetric vacua is in agreement with the Borel-Weil-Bott formula \cite{bbw},
\begin{eqnarray}
H^r(\mathbb P^2, \mathcal T_{\mathbb P^2} (-5)) = \begin{cases} 0 & r = 0 , 1 ; \\ \mathbb C^3 & r = 2 . \end{cases} \nonumber
\end{eqnarray}
As far as the author is aware, there is no direct  gauged linear realisation of an $\mathcal N = 2$ non-linear model whose ground states live in the cohomology groups of twists of the tangent bundle.  The locally free resolution $\mathcal F^\bullet$ provides an indirect realisation of such a sigma model. This technique generalises straightforwardly for arbitrary exterior powers of $\mathcal T_{\mathbb P^N}$.

\paragraph{(iii)}
In our third example, we consider a theory with two chiral multiplets and a single Fermi multiplet of charge $d$, with a degree $d$ superpotential $E(\phi_0, \phi_1, \phi_2)$ with non-degenerate gradient vector $\partial_i E$. We set the Wilson line charge $k$ to zero. In the $\zeta \to \infty$ limit, this theory reduces to an $\mathcal N = 2$ non-linear sigma model on the plane curve $C = \left( E(x_0, x_2, x_2) = 0 \right) \subset \mathbb P^2$. If we integrate out the Fermi multiplets, we obtain a theory with a two-dimensional Chan-Paton space associated to the complex,
\begin{eqnarray}
\mathcal F^\bullet = \mathcal O_{\mathbb P^2}(-d) \overset{ E }{\xrightarrow{\hspace*{0.7cm}}}  \underline{\mathcal O_{\mathbb P^2} }.\nonumber
\end{eqnarray}
Because $H^2(\mathcal O_{\mathbb P^2} (-d))$ is isomorphic to the vector space spanned by degree $d-3$ monomials in $x_0, x_1$ and $x_2$, one learns that $E_1^{p,q} ({\rm NL}) = E_2^{p,q}({\rm NL})$ are as shown below. 
\begin{center}
\begin{picture}(210,68)
\put (68,0) {$0$}
\put (133,0) {$\underline{\mathbb C}$}

\put (68,25) {0}
\put (135,25) {0}

\put (43,50) {$\mathbb C^{(d-1)(d-2)/2}$}
\put (134,50) {0}
\end{picture}
\end{center}
All subsequent differentials vanish, so the spectral sequence terminates at $s = 2$. The number of vacua is
\begin{eqnarray}
\mathcal H^r_0({\rm NL}) = \begin{cases} \mathbb C & r = 0; \\ \mathbb C^{(d-1)(d-2)/2} & r = 1 ;\\ 0 & r \notin \{0, 1 \} . \end{cases} \nonumber
\end{eqnarray}
Since $\mathcal F^\bullet$ is quasi-isomorphic to $i_\star O_C$, where $i : C \hookrightarrow \mathbb P^2$ is the natural embedding, we expect this to coincide with
\begin{eqnarray}
H^r(C, O_C) = \begin{cases} \mathbb C & r = 0; \\ \mathbb C^g & r = 1 ,\end{cases} \nonumber
\end{eqnarray}
where $g$ is the genus of the curve $C$. But the genus-degree formula states that
\begin{eqnarray}
g = \frac 1 2 (d-1)(d-2) ,\nonumber
\end{eqnarray}
so the results are consistent. When we revisit this example in Section 3, we will find that the vacua of $R$-charge $r=0$ and $r=1$ behave differently on the Coulomb branch.

\paragraph{(iv)}
Our final example will be of particular importance for understanding some subtleties of the Coulomb branch description of our $\mathcal N = 2$ theories. Working in $\mathbb P^1$ for simplicity, the complex we wish to consider is a three-term Koszul complex,
\begin{eqnarray}
\mathcal F^\bullet = \mathcal O_{\mathbb P^1}(-2) \overset{\left( \begin{array}{c} x_0 \\ x_1 \end{array} \right) }{\xrightarrow{\hspace*{1.4cm}}} \mathcal O_{\mathbb P^1}(-1)^{\oplus 2}  \overset{\left( \begin{array}{cc} -x_1 & x_0  \end{array} \right) }{\xrightarrow{\hspace*{1.4cm}}} \underline{\mathcal O_{\mathbb P^1}}.\nonumber
\end{eqnarray}
The complex is exact, so it is quasi-isomorphic to zero, and therefore its hypercohomology must vanish in all degrees. Indeed, one way to realise this complex in the gauged linear model is to introduce a pair of Fermi multiplets with superpotentials $E_0 = \phi_0$ and $E_1 = \phi_1$; the Higgs branch is the complete intersection $(x_0 = x_1 = 0) \subset \mathbb P^1$, which is the empty set, so the $\zeta \to \infty$ limit of the theory is governed by a non-linear model with empty target space, and hence, no supersymmetric ground states.

\paragraph{}
Let us examine what must happen in the spectral sequence in order that the hypercohomology groups vanish. By the usual method, we obtain $E_1^{p,q} ({\rm NL}) = E_2^{p,q} ({\rm NL}) $ as shown below:
\begin{center}
\begin{picture}(200,46)
\put (34,0) {0}
\put(100,0) {0}
\put(166,0) {$\underline{\mathbb C}$}

\put (34,33) {$\mathbb C$}
\put(100,33) {0}
\put(166,33) {0}

\put (124, 19) {$d_2^{-2,1}$}
\put (63,28){\vector(4,-1){80} }

\end{picture}
\end{center}
There are two possibilities for the differential $d_2^{-2,1}$: either $d_2^{-2,1}$ is the zero map or it is an isomorphism. We know that $\mathcal F^\bullet$ is quasi-isomorphic to zero, so it must be that $d_2^{-2,1}$ is an isomorphism. This ensures that for all $p$ and $q$,
\begin{eqnarray}
E_\infty^{p,q} = E_3^{p,q} = 0 . \nonumber
\end{eqnarray}
and so, $\mathcal H^r_0 ( {\rm NL}) = 0$ for all $r$, as required. In Section 3.6, we will explain how to interpret this $d_2^{-2,1}$ differential as a tunnelling process on the Coulomb branch.

\section{Supersymmetric Vacua: Coulomb Branch Analysis}

\paragraph{}
Having analysed the spectrum of supersymmetric ground states in the limit $\zeta \to \infty$, we now study the opposite regime, $\zeta \to 0$. In this limit, the Higgs branch of classical vacua shrinks to zero size, and the Coulomb branch emerges. The Coulomb branch is spanned by the vector multiplet scalar $\sigma \in \mathbb R$. For large values of $| \sigma |$, the chiral multiplet fields have large mass and can be integrated out to give an effective action for the vector multiplet fields. The aim of this section is to characterise this effective action on the Coulomb branch, and to determine the fate of the individual supersymmetric ground states that we found in the Higgs branch picture, as we reduce $\zeta$ from infinity to zero. 

\paragraph{}
As we lower $\zeta$, we expect the quantum wavefunctions of the supersymmetric vacua to spread along the Coulomb branch, becoming non-normalisable at infinity when $\zeta = 0$ \cite{denef, moredenef,  hky}. But we would like to understand this process better: in particular, we would like to understand the nature of the wavefunctions for individual states; moreover, we would like to propose a scheme for counting how many states of each $R$-charge escape at $\sigma \to -\infty$, and how many escape at $\sigma \to + \infty$, and interpret these numbers geometrically. The key is to identify a pair of spectral sequences that simplify the analysis of the effective action on the Coulomb branch, and understand how they relate to the spectral sequence $E_s^{p,q} ({\rm NL})$ defined in Section 2.3.

\subsection{Free Theories}

\paragraph{}
We start by reviewing the Coulomb branch analysis of free $\mathcal N = 2$ theories with Wilson lines as carried out in \cite{hky}. This will serve as the basis for the analysis of theories with potentials. The action for these free theories takes the form
\begin{eqnarray}
L &=& \frac 1 {2e^2} ( \dot \sigma ^2 + i \bar \lambda \dot \lambda + D^2 ) - \zeta D \nonumber \\
&& + D_t \bar \phi_i D_t \phi_i +  \bar \phi_i (D - \sigma^2 ) \phi_i + i \bar\psi_i D_t \psi_i + \bar \psi_i  \sigma \psi_i  - i \bar \phi_i \lambda \psi_i + i \bar\psi_i \bar \lambda \phi_i  \nonumber \\
&&  - \left(k+  \frac {N+1} 2 \right)(u_t + \sigma). \nonumber
\end{eqnarray}
We proceed in two steps. First, we fix a particular supersymmetric background for the vector multiplet fields, and quantise the chiral multiplet fields in this background. Next, by examining how the chiral multiplet vacua respond to perturbations to the background, we deduce an effective potential for the vector multiplet fields.

\paragraph{}
For the first step, we fix the supersymmetric background,
\begin{eqnarray}
\sigma = {\rm constant}, \ \ \ \ \   \lambda = 0,  \ \ \ \ \ \  D = 0 , \nonumber
\end{eqnarray} 
and choose the gauge,
\begin{eqnarray}
u_t = 0 .\nonumber
\end{eqnarray}
In this background, the Hamiltonian for the matter fields (i.e. chiral multiplet fields) takes the form,
\begin{eqnarray}
& H = \dot{\bar \phi}_i \dot \phi_i +  \bar\phi_i \sigma^2 \phi_i - \frac 1 2 \sigma [ \bar \psi_i , \psi_i ] + \sigma (k + \frac {N+1} 2). \nonumber
\end{eqnarray}
Meanwhile, the equation of motion for $u_t$ imposes a constraint on the Hilbert space, defining which states are to be considered as physical. The constraint is that the $U(1)$ charge operator, which in $u_t = 0$ gauge takes the form
\begin{eqnarray}
& G = i \bar \phi_i \dot\phi_i - i \dot{\bar \phi}_i \phi_i + \frac 1 2 [ \bar\psi_i , \psi_i ] + (k + \frac{N+1}2) , \nonumber
\end{eqnarray}
vanishes on physical states.

\paragraph{}
Provided that $\sigma \neq 0 $, the matter fields are massive, and one can diagonalise the Hamiltonian and charge operators by introducing creation and annihilation operators,
\begin{eqnarray}
& \phi_i = \frac 1 {\sqrt{2|\sigma|}} (a_i + b_i^\dagger), \ \ \ \ \ \ \ \ \dot \phi_i = - i \sqrt{\frac{|\sigma|}{2}} (a_i -  b_i^\dagger). \nonumber
\end{eqnarray}
The Hamiltonian and charge operators now take the diagonalised form,
\begin{eqnarray}
 H &=&  | \sigma | (a_i^\dagger a_i + b_i^\dagger b_i + N+1) + \sigma (- \bar\psi_i \psi_i + k+ N+1 ), \nonumber \\
 G & = & a_i^\dagger a_i - b_i^\dagger b_i + \bar\psi_i \psi_i + k.
\nonumber
\end{eqnarray}
The number of supersymmetric vacua in this matter theory depends both on the Wilson line charge $k$ and whether the fixed value of $\sigma $ is negative or positive. Let us define a state $| 0 \rangle$ such that
\begin{eqnarray}
a_i | 0 \rangle = b_i | 0 \rangle = \psi_i | 0 \rangle = 0 . \nonumber
\end{eqnarray}
If $\sigma < 0 $, then the space of supersymmetric ground states for the matter sector is
\begin{eqnarray}
& \mathcal H_0 (\sigma < 0 ) = \left\langle (b_0^\dagger)^{n_0} ... (b_N^\dagger)^{n_N} |0 \rangle \ :  \sum_{i=0}^N n_i = k \right\rangle .
\nonumber
\end{eqnarray}
On the other hand, if $\sigma > 0 $, then the space of vacua is
\begin{eqnarray}
&  \mathcal H_0 (\sigma > 0 ) = \left\langle (a_0^\dagger)^{n_0} ... (a_N^\dagger)^{n_N} \bar\psi_0 ... \bar\psi_N|0 \rangle \ :   \sum_{i=0}^N n_i = -  k - N - 1 \right\rangle .
\nonumber
\end{eqnarray}
This completes the first step.

\paragraph{}
Next, we choose any particular ground state of the matter theory and look at how it responds if the vector multiplet background deviates from the supersymmetric background $\sigma = {\rm constant}, \lambda = 0$, $ D = 0$. This will enable us to determine the potential terms in the effective vector multiplet action. In fact, this process must be repeated for every one of the matter theory vacua, giving an effective vector multiplet action for each one of them; together, all these effective actions will represent different sectors of the theory.

\paragraph{}
As the effective Coulomb branch action is constrained by $\mathcal N = 2$ supersymmetry, it suffices to examine the effect of turning on a non-zero vacuum expectation value for $D$. Under this perturbation, the $a_i$ and $b_i$ operators are modified by replacing
\begin{eqnarray}
| \sigma | \mapsto \sqrt{ \sigma^2 - D} \approx | \sigma | - \frac{D}{2| \sigma |}.  \nonumber
\end{eqnarray}
A similar change occurs in front of the $(a_i^\dagger a_i + b_i^\dagger b_i + N+1)$ term in the Hamiltonian operator. The result is a shift in the energy of the ground states:
\begin{eqnarray}
\Delta H = \begin{cases} -  \frac {k + N+1}{2|\sigma|} D, & \ \sigma < 0, \  k \geq 0; \\
\ \ \frac {k}{2\sigma} D, & \ \sigma > 0, \  k \leq -N-1. \end{cases} \nonumber
\end{eqnarray}
Invariance under $\mathcal N = 2$ supersymmetry requires that the effective vector multiplet Lagrangian associated with a given matter ground state is of the form,
\begin{eqnarray}
L = \frac 1 {2e^2} (\dot \sigma^2 + i \bar \lambda \dot \lambda + D^2) + h'(\sigma) D - \frac 1 2 h''(\sigma) \bar \lambda \lambda, \nonumber
\end{eqnarray} 
and in view of how the energy of the ground state responds to fluctuations of $D$, we have
\begin{eqnarray}
h(\sigma) = \begin{cases} - \frac {k+N+1}2  \log | \sigma | + \zeta | \sigma | , & \ \sigma < 0 , \ k \geq 0; \\
\ \ \ \ - \frac k 2 \log \sigma - \zeta \sigma, & \ \sigma > 0 , \ k \leq - N - 1. \end{cases} \nonumber
\end{eqnarray}
With a positive Fayet-Iliopoulos parameter $\zeta > 0 $, one finds by a standard calculation \cite{wittensusy} that for each matter ground state, the effective vector multiplet theory has a single supersymmetric vacuum, with a normalisable wavefunction,
\begin{eqnarray}
\Psi(\sigma) = \begin{cases} e^{-h(\sigma)} | 0 \rangle = | \sigma |^{(k+N+1)/2} e^{- \zeta |\sigma|} |0 \rangle, & \ \sigma < 0 , \ k \geq 0; \\ e^{h(\sigma)}\bar\lambda | 0 \rangle =  \sigma^{-k/2} e^{-\zeta \sigma} \bar\lambda | 0 \rangle, & \  \sigma > 0 , \ k \leq - N - 1. \end{cases}   \nonumber
\end{eqnarray}

\paragraph{}
Notice that the wavefunction is peaked around $ \sigma \sim \zeta^{-1}$. As $\zeta \to \infty$, the support of the wavefunction moves towards $\sigma \to 0$, where the Coulomb branch attaches to the Higgs branch. As $\zeta \to 0$, the support of the wavefunction moves far along the Coulomb branch, towards $\sigma \to - \infty$ or $\sigma \to + \infty$; the wavefunction becomes non-normalisable when $\zeta = 0$.

\paragraph{}
In this analysis, we quantised the chiral multiplets in the background of a fixed vector multiplet profile. The price we pay for this simplification is that we obtain the corrections to the vector multiplet action only at lowest order in the derivative expansion \cite{denef,me3}. On dimensional grounds, higher derivative corrections are negligible for $|\sigma| \gg e^{3/2}$ but become important for $| \sigma | \ll e^{3/2}$. For example, the metric in front of the kinetic term for $\sigma$ may diverge in the limit $\sigma \to 0$, putting the Higgs branch at infinite distance; this sort of behaviour was first observed in $\mathcal N=(4,4)$ theories in two dimensions \cite{conformaltheory}. Because the support of the wavefunctions for our supersymmetric ground states approaches $\sigma = 0$ in the $\zeta \to \infty$ limit, the expressions for the wavefunctions obtained by solving our effective Coulomb branch action are unreliable as $\zeta \to \infty$; one gets a better description of the wavefunctions for these states as $\zeta \to \infty$ from the Higgs branch picture in Section 2.

\paragraph{}
For the purposes of determining whether our ground states are normalisable, one expects that these higher order corrections are unimportant. This is because the only conceivable way for the wavefunctions to escape to infinity is for their support to move far along the non-compact Coulomb branch, where the higher order corrections can be ignored. But wavefunctions cannot escape at $\sigma = 0$, because this is where the Coulomb branch attaches to the Higgs branch, and the Higgs branch is compact in this model.

\paragraph{}
Let us check that this Coulomb branch description provides the correct number of vacuum states for each $R$-charge. Note that each $\bar \psi_i$ carries $R$-charge $+1$ while $\bar \lambda  $ carries $R$-charge $-1$. As mentioned already, there is a different effective Coulomb branch theory corresponding to each ground state of the matter Hamiltonian, and each supplies a single supersymmetric ground state. Supersymmetric ground states associated with matter vacua on the $\sigma < 0 $ part of the Coulomb branch carry total $R$-charge $0$, and those associated with matter vacua on the $\sigma > 0 $ part of the Coulomb branch carry $R$-charge $N$, taking into account the contribution of $-1$ from $\bar\lambda$. So we have
\begin{eqnarray}
\mathcal H_0^r \cong   \begin{cases} \left\langle (b_0^\dagger)^{n_0} ... (b_N^\dagger)^{n_N}  |0 \rangle \ :  \  \sum_{i=0}^N n_i = k \right\rangle,  &  r = 0 ;\\
\left\langle (a_0^\dagger)^{n_0} ... (a_N^\dagger)^{n_N} \bar\psi_0 ... \bar\psi_N  |0 \rangle \ :  \  \sum_{i=0}^N n_i = -  k - N - 1 \right\rangle, & r = N; \\
0 & r \notin \{0 , N\}.   \end{cases} \nonumber
\end{eqnarray}
This agrees with the answer in terms of the cohomology groups $H^r(\mathcal O_{\mathbb P^N}(k))$ from the Higgs branch picture in Section 2, and concludes the review of the Coulomb branch analysis of free $\mathcal N = 2$ theories from \cite{hky}.

\subsection{Theories with Potentials: Matter Sector}

\paragraph{}
We now analyse the effective Coulomb branch description for the general class of theories described in Section 2.1. The action is
\begin{eqnarray}
L &= &\frac 1 {2e^2} ( \dot \sigma ^2 + i \bar \lambda \dot \lambda + D^2 ) - \zeta D \nonumber \\
&& + D_t \bar \phi_i D_t \phi_i +  \bar \phi_i (D - \sigma^2 ) \phi_i + i \bar\psi_i D_t \psi_i + \bar \psi_i  \sigma \psi_i - i \bar \phi_i \lambda \psi_i + i \bar\psi_i \bar \lambda \phi_i  \nonumber \\
&&  - \left(\rho + {\rm diag}\left( \frac {N+1} 2 , ... , \frac {N+1} 2\right)\right)  (u_t + \sigma) -  \{ f, f^\dagger \} + \psi_i \partial_i f -  \bar\psi_i \partial_{\bar i} f^\dagger,
\nonumber
\end{eqnarray}
where, as explained in Section 2.1, $\rho$ and $f$ encode the data for the complex of vector bundles,
\begin{eqnarray}
\mathcal F^\bullet = \dots \to \oplus_{j=1}^{D^{p-1}} \mathcal O_{\mathbb P^N} (k_j^{p-1}) \overset{f^{p-1}} {\rightarrow} \oplus_{j=1}^{D^{p}} \mathcal O_{\mathbb P^N} (k_j^{p}) \overset{f^{p}} {\rightarrow} \oplus_{j=1}^{D^{p+1}} \mathcal O_{\mathbb P^N} (k_j^{p+1})  \to \dots . \nonumber
\end{eqnarray}

\paragraph{}
Once again, we begin by quantising the matter fields (i.e. the chiral multiplet fields) in a fixed supersymmetric background of the form $\sigma = {\rm constant}, \lambda = 0, D = 0$. Since we now have non-linear potential terms, it is impossible to solve for the matter vacua exactly. Nonetheless, we will find that it is still possible to count how many matter vacua there are for any  background value of $\sigma$.

\paragraph{}
Let us recall some general observations about supersymmetric quantum mechanics from \cite{wittensusy}. Because the $R$-charge operator commutes with the Hamiltonian, it is possible to find a simultaneous basis of eigenvectors for the Hamiltonian and the $R$-charge. The action of the supercharge on one of these eigenstates preserves its energy but raises its $R$-charge by one. The supercharge squares to zero, and one obtains a complex,
\begin{eqnarray}
\dots \overset{Q}{\to} \mathcal H^{r-1} \overset{Q}{\to} \mathcal H^{r} \overset{Q}{\to} \mathcal H^{r+1} \overset{Q}{\to} \dots \nonumber
\end{eqnarray}  
This $Q$-complex is exact on states of positive energy. Indeed, if $H | \psi \rangle = E | \psi \rangle$ with $E > 0$, and if $Q | \psi \rangle = 0$, then the supersymmetry algebra $H = \{ Q, \bar Q \} $ immediately gives $| \psi \rangle = Q\left(E^{-1} \bar Q  | \psi \rangle \right)$. However, the action of $Q$ is trivial on states of zero energy. Hence the space of supersymmetric vacua of a given $R$-charge $r$ is isomorphic to the homology of the $Q$-complex at grading $r$.
\begin{eqnarray}
\mathcal H^r_0 = H_Q^r (\mathcal H^\bullet ).  \nonumber
\end{eqnarray}
Of course, this statement merely tells us how many supersymmetric vacua there are and which $Q$-cohomology classes they lie in; it does not tell us their wavefunctions.

\paragraph{}
Let us apply this to our matter theory in a fixed background, $\sigma = {\rm constant}, \lambda = 0, D = 0$. We have a pair of supercharges,
\begin{eqnarray}
Q = i \bar\psi_i \dot\phi_i - \sigma \bar\psi_i \phi_i +  f(\phi_i), \ \ \ \ \ \ \ \ \ \ \ 
\bar Q = - i \dot{\bar\phi}_i \psi_i - \sigma \bar\phi_i \psi_i +  f^\dagger (\bar \phi_i), \nonumber
\end{eqnarray}
and in principle we can count the number of supersymmetric vacua by taking homology in either $Q$ or $\bar Q$. As it turns out,  the analysis works differently depending on whether the background value of $\sigma$ we fix is positive or negative. We describe each of these cases in turn.

\subsubsection*{Ground States of the Matter Hamiltonian for negative $\sigma$}

\paragraph{}
When $\sigma < 0$, the expressions for the supercharges in terms of the creation and annihilation operators defined in Section 3.1 are
\begin{eqnarray}
Q & = & \sqrt{2|\sigma |} \bar\psi_i a_i +  f\left( \frac {1}{\sqrt{2|\sigma|}} (a_i + b_i^\dagger) \right), \nonumber \\
\bar Q & = & \sqrt{2| \sigma |} a_i^\dagger \psi_i + f^\dagger \left( \frac 1 {\sqrt{2| \sigma |}} (b_i + a_i^\dagger) \right). \nonumber
\end{eqnarray}
We choose to take homology in the $Q$ operator; the advantage in making this choice will become apparent in due course. We write the Hilbert space for the matter sector as $\mathcal H (\sigma < 0 )$, appending the symbol $(\sigma < 0 ) $ as a reminder that our chosen fixed background value for $\sigma$ is negative, and we decompose the Hilbert space as a direct sum of the spaces $\mathcal H^{p,q} (\sigma < 0 )$ spanned by states of Chan-Paton grading $p$ and $\bar\psi_i$ particle number $q$.
\begin{eqnarray}
\mathcal H (\sigma < 0 )= \underset{p,q}{\oplus} \mathcal H^{p,q} (\sigma < 0 ) .\nonumber
\end{eqnarray}
 Since the Chan-Paton grading contributes to the $R$-charge of a state, and $\bar\psi$ carries $R$-charge $+1$, the space of states of $R$-charge $\hat r$ are
\begin{eqnarray}
\mathcal H^{\hat r} (\sigma < 0) = \underset{p + q = \hat r} \oplus \mathcal H^{p,q} (\sigma <0 ). \nonumber
\end{eqnarray}
The reason for denoting the $R$-charge as $\hat r$, rather than as $r$, is that the $R$-charge $ \hat r$ takes into account contributions from the Chan-Paton grading and from the $\bar\psi_i$ particle number, but not from the $\bar \lambda $ particle number. We will reserve use of the symbol $r$ for the full $R$-charge.

\paragraph{}
We define the pair of differentials,
\begin{eqnarray}
& \delta^{p,q} : \mathcal H^{p,q} (\sigma <  0)  \to \mathcal H^{p, q+1} (\sigma < 0 ), \ \ \ \ \ \ \ & \delta^{p,q} = \sqrt{2|\sigma|} \bar\psi_i a_i \nonumber, \\
&{\delta'}^{p,q} : \mathcal H^{p,q}(\sigma < 0 ) \to \mathcal H^{p+1, q} (\sigma < 0 ), \ \ \ \ \ \  & {\delta'}^{p,q} = (-1)^q f^p \left( \frac{1}{\sqrt{2 | \sigma |}} (a_i + b_i^\dagger) \right), \nonumber
\end{eqnarray}
that square to zero and anticommute with one another, such that the supercharge operator is their sum,
\begin{eqnarray}
Q^{p,q} = \delta^{p,q} + {\delta'}^{p,q}. \nonumber
\end{eqnarray}
Note that we have included a factor $(-1)^q$ in the definition of $\delta ' $, coming from the fact that the endomorphism $f$ anticommutes with fermions.  As explained above, the space of supersymmetric ground states of the matter Hamiltonian for our fixed value of $\sigma$ is isomorphic to the $Q$-cohomology,
\begin{eqnarray}
\mathcal H_0^{\hat r} (\sigma < 0) \cong H^{\hat r}_Q \left(\mathcal H^\bullet (\sigma <0 ) \right) . \nonumber
\end{eqnarray}
We can compute this $Q$-cohomology using a spectral sequence for the bigraded complex $\mathcal H^{p,q} (\sigma < 0 )$. We call this spectral sequence $E_s^{p,q}(\sigma <0)$ to avoid confusion with the spectral sequence $E_s^{p,q}({\rm NL})$ defined in Section 2.3.

\paragraph{}
As usual, $E_0^{p,q} (\sigma < 0)  = \mathcal H^{p,q}(\sigma < 0) $. To find $E_1^{p,q} (\sigma < 0 )$, we must take the homology in  $d_0^{p,q} = \delta^{p,q}$. Observe that $\delta^{p,q}$ preserves both the Chan-Paton state and the $b_i^\dagger$ particle content, so for the purposes of finding the $\delta^{p,q}$ holomology, we may as well restrict our attention to a sector of the Hilbert space with a fixed Chan-Paton factor and a fixed $b_i^\dagger$ particle content. Furthermore, the $U(1)$ charge operator $G$ commutes with $Q$, so we can safely postpone imposing the gauge fixing condition $G = 0$ till the end. A generic state in any such sector takes the form
\begin{eqnarray}
 \left( \sqrt{2|\sigma |} a_0^{\dagger} \right)^{m_0} \dots  \left( \sqrt{2|\sigma |} a_N^{\dagger} \right)^{m_N} \bar\psi_{i_1} \dots  \bar\psi_{i_q} | b^\dagger, {\rm CP} \rangle,  \nonumber
\end{eqnarray}
where $m_0, ... , m_N \geq 0$ and $ i_1, ... , i_q \in \{ 0, ... , N \}$. By identifying
\begin{eqnarray}
\sqrt{2| \sigma |} a_i^\dagger \mapsto z_i, \ \ \ \ \ \ \ \ \ \bar\psi_i \mapsto dz_i,
\nonumber
\end{eqnarray}
we see that there is a correspondence between the quantum states in this sector and holomorphic forms on  $\mathbb C^{N+1}$,
\begin{eqnarray}
z_0^{m_0} \dots z_N^{m_N} dz_{i_1} \wedge \dots \wedge dz_{i_q} \in \Omega^{q,0} (\mathbb C^{N+1} ) .
\nonumber
\end{eqnarray}
Under this correspondence, the action of the differential $d_0^{p,q} = \delta^{p,q}$ is the action of the Dolbeault operator on $\mathbb C^{N+1}$,
\begin{eqnarray}
d_0^{p,q} \ \mapsto \ \partial . \nonumber
\end{eqnarray}
At this point, the advantage in choosing $Q$ rather than $\bar Q$ is apparent: the homology of $E_0^{p,q}$ in $d_0^{p,q}$ follows directly from the Poincar\'{e} lemma,
\begin{eqnarray}
H_{\partial}^{q,0} (\mathbb C^{N+1} ) = \begin{cases} \mathbb C & q = 0; \\ 0 & q \geq 1. \end{cases} \nonumber
\end{eqnarray}
We learn that the $E_1^{p,q}(\sigma < 0 )$ terms in our spectral sequence are only non-trivial in the row $q = 0$, and are given by
\begin{eqnarray}
& E_1^{p,0} (\sigma < 0 ) \cong 
\oplus_{j=1}^{D^p} \left\langle  \left( \frac 1 {\sqrt{2 | \sigma |}} b_0^\dagger \right)^{n_0} ... \left( \frac 1 {\sqrt{2 | \sigma |}} b_N^\dagger\right)^{n_N} | k_j^p \rangle \ \ : \ \sum_{i=0}^N n_i = k_j^p  \right\rangle ,
\nonumber
\end{eqnarray}
\begin{eqnarray}
E_1^{p,q} (\sigma < 0 ) = 0, \ \ \ \ \ \ \ \ \ \  \ \ \ \ \   1 \leq q \leq  N+1. \nonumber
\end{eqnarray}
(The states $\{ | k_j^p \rangle \} $ are defined to span the ground states in the Chan-Paton sector $\mathcal V^p$.)  The condition $\sum_{i=0}^N n_i = k_j^p $ comes from finally imposing the gauge fixing constaint,
\begin{eqnarray}
&  G   = a_i^\dagger a_i - b_i^\dagger b_i + \bar\psi_i \psi_i  + \rho = 0.
\nonumber
\end{eqnarray}
The expression for $E_1^{p,0} (\sigma < 0 )$ is similar to something we have already encountered in Section 2.3: if we make the replacement,
\begin{eqnarray}
& \frac {1} { \sqrt{2|\sigma|} } b_i^\dagger \ \mapsto \ x_i , \nonumber 
\end{eqnarray}
then the expression for $E_1^{p,0} (\sigma < 0 )$ coincides with $H^0(\mathcal F^p) = \oplus_{j = 1}^{D^p} H^0 \left(\mathcal O_{\mathbb P^N}(k_j^p)\right) $. So we have
\begin{eqnarray}
E_1^{p, q } (\sigma < 0 ) \cong \begin{cases} E_1^{p,0} ({\rm NL})  & q = 0 ; \\  \ \ \ \ \ 0 & 1 \leq q \leq N + 1 .\end{cases} \nonumber
\end{eqnarray}

\paragraph{}
Next, we take homology in $d_1^{p,q} = {\delta ' }^{p,q} $ to compute $E_2^{p,q}(\sigma < 0 )$. Clearly, $E_2^{p,q}(\sigma < 0)$ is automatically zero for $1 \leq q \leq N+1 $, and we only have to deal with $E_2^{p,0} (\sigma < 0 )$. We note that the representatives of the $d_0$-cohomology classes spanning $E_1^{p,0}$, as written above, contain no $a_i^\dagger$ operators, so they are annihilated by $a_i$. The action of ${\delta'}^{p,0}$ on these classes is then simply
\begin{eqnarray}
& {\delta'}^{p,0} =  f^p \left( \frac 1 {\sqrt{2|\sigma |}} b_i^\dagger \right). \nonumber
\end{eqnarray} 
If we make the replacement $\frac {1} { \sqrt{2|\sigma|} } b_i^\dagger \ \mapsto \ x_i $, we observe that this is the group homomorphism, $d_1^{p,q}  = f^p (x_i) : E_1^{p,q} ({\rm NL} ) \to E_1^{p+1, q} ({\rm NL}) $ that appeared in the spectral sequence for the non-linear model in Section 2.3. So we have
\begin{eqnarray}
E_2^{p, q } (\sigma < 0 ) \cong \begin{cases} E_2^{p,0} ({\rm NL})  & q = 0; \\ \ \ \ \ \ 0 & 1 \leq q \leq N + 1 .\end{cases} \nonumber
\end{eqnarray}
Since $E_2^{p,q}(\sigma < 0 ) $ vanishes for $1 \leq q \leq N+1 $, the spectral sequence terminates here.
\begin{eqnarray}
E_\infty^{p,q} (\sigma < 0 ) \cong \begin {cases} E_2^{p,0} ({\rm NL}) &  q = 0; \\ \ \ \ \ \  0 & 1 \leq q \leq N + 1 . \end{cases} \nonumber
\end{eqnarray}
We conclude that the number of ground states of the matter Hamiltonian, in a fixed vector multiplet background with $\sigma  < 0 $, is
\begin{eqnarray}
\mathcal H^{\hat r}_0 (\sigma < 0 ) \cong \underset{p+q = \hat r} \oplus E_\infty^{p, q } (\sigma < 0 ) \cong E_2^{\hat r,0} ({\rm NL}). \nonumber
\end{eqnarray}

\subsubsection*{Ground States of the Matter Hamiltonian for positive $\sigma$}

\paragraph{}
We repeat our analysis in the case where the fixed value of $\sigma$ lies on the positive half of the Coulomb branch. Since the expressions for the creation and annihilation operators depend on $| \sigma |$ but the supercharges depend on $\sigma$, the form of $Q$ and $\bar Q$ when expressed in terms of $a_i$ and $b_i$ is different, now that $\sigma$ is positive.
\begin{eqnarray}
Q & = & - \sqrt{2\sigma } \bar\psi_i b_i^\dagger +  f\left( \frac {1}{\sqrt{2 \sigma }} (a_i + b_i^\dagger) \right) , \nonumber \\
\bar Q & = & -  \sqrt{2 \sigma } b_i  \psi_i +  f^\dagger \left( \frac 1 {\sqrt{2  \sigma }} (b_i + a_i^\dagger) \right) . \nonumber
\end{eqnarray}
This time, it is more convenient to count supersymmetric vacua in the matter sector by taking homology in $\bar Q$ rather than $Q$, as this will allow us to use the same trick with the Poincar\'{e} lemma. But choosing $\bar Q$ means that the differentials on the bigraded complex $\mathcal H^{p,q} (\sigma > 0 )$ point opposite to the conventional direction.
\begin{eqnarray}
& \delta^{p,q} : \mathcal H^{p,q} (\sigma >  0)  \to \mathcal H^{p, q-1} (\sigma > 0 ), \ \ \ \ \ \ \ & \delta^{p,q} = - \sqrt{2|\sigma|} b_i \psi_i, \nonumber \\
&{\delta'}^{p,q} : \mathcal H^{p,q}(\sigma > 0 ) \to \mathcal H^{p-1, q} (\sigma > 0 ), \ \ \ \ \ \  & {\delta'}^{p,q} = (-1)^q  (f^p)^\dagger \left( \frac{1}{\sqrt{2 | \sigma |}} (b_i  + a_i^\dagger) \right). \nonumber
\end{eqnarray}

\paragraph{}
We denote the spectral sequence associated to this bigraded complex as $E_s^{p,q}(\sigma > 0 )$. The states in a sector of fixed Chan-Paton factor and fixed $a_i^\dagger$ number take the form
\begin{eqnarray}
\left( - \sqrt{2\sigma} b_0^\dagger \right)^{m_0} \dots \left( - \sqrt{2 \sigma} b_N^\dagger \right)^{m_N} \psi_{i_1} \dots \psi_{i_{N+1-q}} \left( \bar\psi_0 \dots  \bar\psi_N | a^\dagger , CP \rangle \right). \nonumber
\end{eqnarray}
By identifying
\begin{eqnarray}
-\sqrt{2\sigma} b_i^\dagger \ \mapsto z_i, \ \ \ \ \ \ \ \ \ \psi_i \mapsto dz_i, \nonumber
\end{eqnarray}
the state becomes
\begin{eqnarray}
z_0^{m_0} \dots  z_N^{m_N} dz_{i_1} \wedge \dots \wedge dz_{i_{N+1-q}} \in \Omega^{N+1-q,0} (\mathbb C^{N+1} ).  \nonumber 
\end{eqnarray}
The differential $d_0^{p,q} = \delta^{p,q}$ acts as
\begin{eqnarray}
d_0^{p,q} = \partial. \nonumber
\end{eqnarray}
Applying the Poincar\'{e} lemma on $\mathbb C^{N+1}$ to compute the homology and imposing the gauge fixing constraint, we obtain
\begin{eqnarray}
& E_1^{p, N+1} (\sigma > 0) = \oplus_{j = 1}^{D^p} \left\langle \left(  \frac 1 {\sqrt{2\sigma}} a_0^\dagger  \right)^{n_0 } ... \left( \frac 1 {\sqrt{2\sigma}} a_N^{\dagger} \right)^{n_N} \bar\psi_0 \dots \bar\psi_N |k_j^p\rangle \   : \ \sum_{i=0}^{N} n_i = - k_j^p - N - 1 \right\rangle,  \nonumber
\end{eqnarray}
\begin{eqnarray}
E_1^{p,q} (\sigma > 0 ) = 0, \ \ \ \ \ \ \ \ \ \  \ \ \ \ \  \   0 \leq q \leq  N. \nonumber
\end{eqnarray}
Notice that this time round, it is the $q = N+1$ row of the spectral sequence that is non-trivial, and the representatives of the homology classes now have $\bar\psi_i$ particle number $N+1$. Comparing this with the expressions for $H^N(\mathcal F^p) = \oplus_{j = 1}^{D^p} H^N \left(\mathcal O_{\mathbb P^N}(k_j^p)\right) $ after applying Serre duality, we see that
\begin{eqnarray}
E_1^{p,q} (\sigma > 0 ) \cong \begin{cases} \ \ \ \ \ \ \  0 & 0 \leq q \leq N; \\  \left( E_1^{p,N} ({\rm NL}) \right)^\vee & q = N+1. \end{cases}  \nonumber
\end{eqnarray}

\paragraph{}
Now we observe that the representatives for the $d_0$-cohomology classes in $E_1^{p,N+1}$ contain no $b_i^\dagger$ operators, so they are annihilated by $b_i$. The action of   ${\delta'}^{p,N+1} $ on these representatives is
\begin{eqnarray}
& {\delta'}^{p,N+1} = (-1)^{N+1}  (f^p)^\dagger \left( \frac {1}{\sqrt{2 \sigma}} a_i^\dagger \right),
\nonumber
\end{eqnarray}
which can be recognised, up to a sign, as $\left( d_1^{p,N} ({\rm NL} )\right)^\vee$; the fact that the differentials in $E_s^{p,q} (\sigma < 0 )$ point opposite to the conventional direction is compensated for by the fact that we are comparing them to the duals of the differentials in $E_s^{p,q}({\rm NL})$. So we obtain
\begin{eqnarray}
E_2^{p, q} (\sigma > 0 ) \cong  \begin{cases} \ \ \ \ \ \ \  0 & 0 \leq q \leq N; \\ \left( E_2^{p,N} ({\rm NL}) \right)^\vee  & q = N+1. \end{cases} \nonumber
\end{eqnarray}
Since $E_2^{p,q}(\sigma > 0 )$ vanishes in the rows $ 0 \leq q \leq N$, the spectral sequence terminates, and
\begin{eqnarray}
E_\infty^{p, q} (\sigma > 0 ) \cong  \begin{cases} \ \ \ \ \ \ \ 0 & 0 \leq q \leq N; \\ \left( E_2^{p, N}({\rm NL}) \right)^\vee  & q = N+1. \end{cases} \nonumber
\end{eqnarray}
Hence the space of supersymmetric ground states of the matter Hamiltonian in the background of a fixed positive $\sigma $ is
\begin{eqnarray}
\mathcal H_0^{\hat r} (\sigma > 0 ) \cong \underset{p + q = \hat r} \oplus E_{\infty}^{p , q}(\sigma > 0 ) \cong \left( E_2^{\hat r-N-1, N } ({\rm NL}) \right)^\vee. \nonumber
\end{eqnarray}

\subsection{Effective Coulomb Branch Action: Complexes of Length $\leq N$}

\paragraph{}
The next step is to characterise the effective action for the vector multiplet variables $(\sigma, \lambda, D)$. Of course, it is impossible to derive an explicit expression for the potential, since we do not have exact expressions for the matter ground states, only their $Q$ or $\bar Q$ cohomology classes and their $R$-charges. Despite this, it is possible to put together a consistent picture of what this effective action should look like, and how the supersymmetric vacua should behave on the Coulomb branch for small $\zeta$.

\paragraph{}
We first treat the simpler case where the range in the Chan-Paton charges is bounded by $N$, the dimension of the projective space.
\begin{eqnarray}
p_{\rm max} - p_{\rm min} \leq N . \nonumber
\end{eqnarray}
In this case, the complex $\mathcal F^\bullet$ contains no more than $N+1$ terms. All differentials in the non-linear model spectral sequence $E_s^{p,q} ({\rm NL}) $ vanish beyond $d_1^{p,q}$,  so $E_s^{p,q} ({\rm NL}) $ terminates at $s = 2$.
\begin{eqnarray}
E_\infty^{p,q} ({\rm NL}) = E_2^{p,q} ({\rm NL}) . \nonumber
\end{eqnarray}

\paragraph{}
Let us consider the Coulomb branch description of such a theory. First, we pick an element of $\mathcal H_0^{\hat r} (\sigma < 0 )$, for some value of $\hat r$. These exist for $\hat r$ lying in the range,
\begin{eqnarray}
\hat r = p_{\rm min} , ... , p_{\rm max}.  \nonumber
\end{eqnarray}
Every such element corresponds to a ground state of the matter Hamiltonian for negative values of $\sigma$. Associated with our chosen element of $\mathcal H_0^{\hat r} (\sigma < 0 )$ is an $\mathcal N = 2$ invariant effective Lagrangian for the vector multiplet. Invariance under $\mathcal N = 2$ supersymmetry implies that it takes the usual form at lowest order in the derivative expansion,
\begin{eqnarray}
L = \frac 1 {2e^2} (\dot \sigma^2 + i \bar \lambda \dot \lambda + D^2) + h'(\sigma) D - \frac 1 2 h''(\sigma) \bar \lambda \lambda. \nonumber
\end{eqnarray} 
Here, the variable $\sigma$ is taken to vary over the interval $\sigma \in (- \infty , 0)$.

\paragraph{}
We examine how the potential behaves in various limits. First, when $\sigma \to - \infty$, the matter Hamiltonian of Section 3.2 reduces to the free theory matter Hamiltonian of Section 3.1, and so the ground states of the matter Hamiltonian reduce to ground states in the free theory. In the free theory, we saw that the dominant contribution to the potential function $h(\sigma)$ as $\sigma \to - \infty$ is the $ \zeta | \sigma |$ term coming from the Fayet-Iliopoulos parameter. So we expect the $h(\sigma)$ function for the interacting theory to behave in the same way at large $| \sigma |$, 
\begin{eqnarray}
h(\sigma) \sim  \zeta | \sigma |,  \ \ \ \ \  \ \ \ \ \ \ \ \ \sigma \to - \infty .\nonumber
\end{eqnarray}
Next, we consider what happens as $\sigma$ tends to $ 0$ from below. As $\sigma \to 0$, the ground state of the matter Hamiltonian disappears from the spectrum, so we expect to find a singularity in the potential function $h (\sigma)$,
\begin{eqnarray}
h(\sigma) \to  \infty, \ \ \ \ \ \ \ \ \ \ \ \ \ \  \sigma \to 0^- .\nonumber
\end{eqnarray}
(The sign of the infinity is chosen to match the sign of the infinity in the free theory.) This suggests that the effective vector multiplet action corresponding to our chosen element of $\mathcal H_0^{\hat r} (\sigma < 0 )$ has one normalisable supersymmetric ground state, namely,
\begin{eqnarray}
\Psi(\sigma) =  e^{-h(\sigma)} | 0 \rangle,  \ \ \ \ \ \ \  \ \ \ \ \sigma < 0  . \nonumber
\end{eqnarray}
This state has $R$-charge $\hat r$ coming from its Chan-Paton grading and $\bar\psi_i$ particle content, and it has zero $\bar \lambda$ number, so its total $R$-charge is $r = \hat r$.  

\paragraph{}
In the $\zeta \to 0$ limit, the wavefunction of the state becomes non-normalisable at the $\sigma \to - \infty$ end of the Coulomb branch. In the $\zeta \to \infty$ limit, the support of the wavefunction penetrates the region $| \sigma | \ll e^{3/2}$ where the Coulomb branch merges with the Higgs branch. This is the region where higher derivative corrections to the effective Lagrangian become important; however, the Higgs branch is compact in all of the models we study, so there is no chance of the wavefunction escaping to infinity on the Higgs branch, and therefore, we expect that there is no issue with normalisability at $\sigma = 0$, despite the possibility of higher derivative corrections.

\paragraph{}
Next, we consider the effective Coulomb branch theory corresponding to an element of $\mathcal H_0^{\hat r} (\sigma > 0 )$. Such elements exist for $\hat r $ in the range,
\begin{eqnarray}
\hat r = p_{\rm min} + N + 1 ,\  \dots, \  p_{\rm \max} + N + 1.  \nonumber
\end{eqnarray}
Applying the same arguments, one expects that corresponding to each such element of $\mathcal H_0^{\hat r}(\sigma > 0)$ there is an effective Coulomb branch theory, now with $\sigma$ lying in the interval $\sigma \in (0, \infty) $, and the potential function $h(\sigma)$ receives its dominant contribution from the Fayet-Iliopoulos parameter at infinity,
\begin{eqnarray}
h(\sigma) \sim  - \zeta \sigma,  \ \ \ \ \  \ \ \ \ \ \ \ \ \sigma \to + \infty. \nonumber
\end{eqnarray}
and has a divergence at $\sigma = 0$,
\begin{eqnarray}
h(\sigma) \to  - \infty, \ \ \ \ \ \ \ \ \ \ \ \ \ \  \sigma \to 0^+. \nonumber
\end{eqnarray}
This suggests that, associated with our chosen element of $\mathcal H_0^{\hat r} (\sigma > 0)$, there is one normalisable supersymmetric ground state, with wavefunction
\begin{eqnarray}
\Psi(\sigma) =  e^{h(\sigma)} \bar \lambda| 0 \rangle,  \ \ \ \ \ \ \  \ \ \ \ \sigma > 0  .\nonumber
\end{eqnarray}
The support of this wavefunction tends to $\sigma \to + \infty$ in the limit $\zeta \to 0 $, and it moves towards the Higgs branch at $\sigma = 0$ in the limit $\zeta \to \infty$. Since this state has $\bar \lambda$ number $1$, its total $R$-charge is $r = \hat r - 1$.

\paragraph{}
Let us compare the number of supersymmetric vacua obtained in our Coulomb branch analysis to the number obtained in the $\zeta \to \infty$ Higgs branch analysis in Section 2.3. For a given total $R$-charge $r$, the number of vacua obtained from the non-linear model on the Higgs branch is
\begin{eqnarray}
\mathcal H_0^r ({\rm NL})  \cong & E_\infty^{ r , 0 } ({\rm NL}) \oplus E_\infty^{ r - N, N} ({\rm NL})  \cong  E_2^{ r , 0 } ({\rm NL}) \oplus E_2^{ r - N, N} ({\rm NL}). \nonumber
\end{eqnarray}
We know that
\begin{eqnarray}
\mathcal H_0^{\hat r} (\sigma < 0 ) & \cong & E_2^{\hat r , 0 } ({\rm NL}), \nonumber \\
\mathcal H_0^{\hat r} (\sigma > 0) & \cong & E_2^{\hat r - N - 1, N} ({\rm NL}).  \nonumber
\end{eqnarray}
(From now on, we drop the dualising symbol from $\left( E_2^{\hat r - N - 1, N} ({\rm NL}) \right)^\vee$ because we only care about the dimensions of these vector spaces.)  So we see that
\begin{eqnarray}
\mathcal H_0^r ({\rm NL}) \cong  \mathcal H_0^{ \hat r = r} (\sigma < 0 ) \oplus \mathcal H_0^{\hat r =  r +1} (\sigma > 0 ). \nonumber
\end{eqnarray}
But according to our Coulomb branch picture, the supersymmetric vacua of $R$-charge $r$ are precisely in correspondence with the elements of $ \mathcal H_0^{\hat  r = r} (\sigma < 0 ) \oplus \mathcal H_0^{\hat r =  r +1} (\sigma > 0 )$, and in the limit $\zeta \to \infty$, all of these vacua migrate towards the Higgs branch. This suggests that all of the supersymmetric vacua that exist in the $\zeta \to \infty$ limit remain as supersymmetric vacua of the theory when we lower $\zeta$, and all are visible as vacua in the effective theory on the Coulomb branch.

\paragraph{}
We can say more: there are two types of supersymmetric ground states of $R$-charge $r$. The ground states represented by elements of $E_2^{r,0}({\rm NL}) \cong \mathcal H_0^{\hat r = r} (\sigma < 0 )$ are supported in the $\sigma < 0$ region for small $\zeta$, and disappear at the $\sigma \to - \infty$ end of the Coulomb branch as $\zeta \to 0$. The ground states represented by elements of $E_2^{r-N,N} ({\rm NL}) \cong \mathcal H_0^{\hat r = r+ 1} (\sigma > 0 )$ are different: they are supported in the $\sigma > 0 $ region for small $\zeta$ and disappear at the $\sigma \to + \infty$ end of the Coulomb branch as $\zeta \to 0$.

\paragraph{}
Let us illustrate this with some examples from Section 2.4. First, we consider example (i), the theory associated to the complex,
\begin{eqnarray}
\mathcal F^\bullet = \mathcal O_{\mathbb P^2}^{\oplus 2} \overset{{\left( \begin{array}{cc}x_0 & x_1 \\ x_1 & x_2  \end{array}\right) }}{\xrightarrow{\hspace*{1.4cm}}}  \underline{\mathcal O_{\mathbb P^2}(1)}^{\oplus 2}.\nonumber
\end{eqnarray}
We interpreted this as a theory that reduces in the $\zeta \to \infty$ limit to a non-linear sigma model on the quadric $\mathbb P^1 \cong (x_0 x_2 - x_1^2 = 0 ) \subset \mathbb P^2$ with three units of Dirac monopole flux through the $\mathbb P^1$. The complex is of length $d_{\rm max} - d_{\rm min} = 1$, which is certainly no greater than the dimension of $\mathbb P^2$, so we can apply the analysis above. In Section 2.4, we computed the spectral sequence for the non-linear model, and found that the only non-trivial $s = 2$ term is $E_2^{0,0}({\rm NL}) = \mathbb C^4$. So this theory has four supersymmetric vacua of $R$-charge $4$, and all four of these vacua disappear at the $\sigma \to - \infty$ end of the Coulomb branch as $\zeta \to 0$.

\paragraph{}
Next, we look at example (ii). The theory is associated to the complex,
\begin{eqnarray}
\mathcal F^\bullet = \mathcal O_{\mathbb P^2}(-5) \overset{{\left( \begin{array}{c} x_0 \\ x_1 \\ x_2  \end{array}\right) }}{\xrightarrow{\hspace*{1.2cm}}}  \underline{\mathcal O_{\mathbb P^2}(-4)}^{\oplus 3}. \nonumber
\end{eqnarray}
It reduces in the $\zeta \to \infty$ limit to a non-linear sigma model in which the ground states are given by the cohomology of the twisted tangent bundle $\mathcal T_{\mathbb P^2}(-5)$. The only non-trivial $s = 2$ term in the non-linear model spectral sequence is $E_2^{0,2} ( {\rm NL} ) = \mathbb C^3$. So the theory has three vacua of $R$-charge $2$, and all three of them disappear at the $\sigma \to + \infty$ end of the Coulomb branch as $\zeta \to 0$.

\paragraph{}
Example (iii) is more interesting. The complex is
\begin{eqnarray}
\mathcal F^\bullet = \mathcal O_{\mathbb P^2}(-d) \overset{ E }{\xrightarrow{\hspace*{0.7cm}}}  \underline{\mathcal O_{\mathbb P^2} },\nonumber
\end{eqnarray}
and the theory reduces to a non-linear model on a plane curve of genus $g = \frac 1 2 (d-1)(d-2)$ as $\zeta \to \infty$. The non-linear sigma spectral sequence has two non-trivial $s = 2$ terms, namely, $E_2^{0,0} = \mathbb C$ and $E_2^{-1,2} = \mathbb C^{g}$, so the non-linear model has a single supersymmetric vacuum of $R$-charge zero and $g$ supersymmetric vacua of $R$-charge one. The Coulomb branch picture suggests that as we reduce $\zeta$, all these states remain as vacua\footnote{The proposed Coulomb branch picture suggests that the supersymmetric vacua in our class of $\mathcal N = 2$ theories are more robust than one might expect based on Witten index arguments alone; a similar point was made for $\mathcal N = 4$ quiver theories in \cite{sen}.}; moreover, the single state of $R$-charge zero moves towards the $\sigma \to - \infty$ end of the Coulomb branch, while the $g$ states of $R$-charge one move towards the $\sigma \to + \infty$ end of the Coulomb branch.

\paragraph{}
Example (iv), the Koszul complex on $\mathbb P^1$, is a complex of length greater than the dimension of $\mathbb P^1$, so it is not covered by the analysis in this section. We will analyse situations of this kind in the next section, and use this particular example as illustration in Section 3.6.

\subsection{Effective Coulomb  Branch Action: Complexes of Arbitrary Length}

\paragraph{}
We now study the situation where the range of Chan-Paton charges is greater than the dimension of the $\mathbb P^N$ target space,
\begin{eqnarray}
p_{\rm max} - p_{\rm min} \geq N+1. \nonumber
\end{eqnarray}
In this case, the Coulomb branch spectral sequences impose tight constraints on the possible numbers of supersymmetric vacua of given $R$-charge localised on each of the two sides of the Coulomb branch, but they do not lead to a unique possibility. However, we will see that if we make the natural assumption that the number of Higgs and Coulomb branch vacua agree (as was the case for complexes of length $\leq N$ in Section 3.3) then we do arrive at a unique possibility for the number of supersymmetric vacua of each $R$-charge localised on each of the two Coulomb branches. It should be stressed that the agreement between the number of Higgs and Coulomb branch vacua for complexes of length $\geq N+ 1$ is a conjecture, and should be considered a part of the proposal rather than a result derived from first principles. (In Section 3.6 we will study an example where this assumption can be verified explicitly.)

\paragraph{}
The root of the difficulties in the case $p_{\rm max} - p_{\rm min} \geq N+1$ is that the spectral sequence $E_s^{p,q} ({\rm NL})$ does not necessarily terminate at $E_2^{p,q}$. Although $E_2^{p,q}({\rm NL}) = ... = E_{N+1}^{p,q} ({\rm NL})$,  it is possible to have non-trivial differentials,
\begin{eqnarray}
d_{N+1}^{\hat r - N - 1, N}({\rm NL}) : E_2^{\hat r - N-1, N} ({\rm NL}) \to E_2^{\hat r , 0 } ({\rm NL}) \nonumber
\end{eqnarray}
for $ \hat r$ in the range,
\begin{eqnarray}
\hat r = p_{\rm min} + N+1 \ , \  ... \ , \  p_{\rm max} . \nonumber
\end{eqnarray}

\paragraph{}
\begin{center}
\begin{picture}(300,70)
\put (0,6) {$E_2^{p_{\rm min}, 0}$}
\put (60,6){$\dots$}
\put (100, 6){$E_2^{p_{\rm min} + N+1, 0}$}
\put (200,6){$\dots$}
\put(290,6) {$E_2^{p_{\rm max}, 0}$}

\put (0,64) {$E_2^{p_{\rm min}, N}$}
\put (100,64){$\dots$}
\put (190, 64){$E_2^{p_{\rm max} - N-1, N}$}
\put (260,64){$\dots$}
\put(290,64) {$E_2^{p_{\rm max}, N}$}

\put (65, 40) {$d_{N+1}^{p_{\rm min},N}$}
\put (18,59){\vector(2,-1){80} }

\put(145, 40) {$ \dots$}

\put (255, 40) {$d_{N+1}^{p_{\rm max}- N-1,N}$}
\put (208,59){\vector(2,-1){80} }

\end{picture}
\end{center}
There are no further non-trivial differentials after $d_{N+1}^{p,q}$, so $E_{N+1}^{p,q} ({\rm NL}) = E_\infty^{p,q} ({\rm NL}) $.

\paragraph{}
The Coulomb branch analysis in Section 3.3 cannot possibly be the whole story, because it gives ${\rm dim } \ E_2^{r, 0 } ({\rm NL})+ {\rm dim} \   E_2^{r-N, N}({\rm NL}) $ supersymmetric vacua of total $R$-charge $r$. This is greater than ${\rm dim } \ E_{N+2}^{r, 0 } ({\rm NL}) + {\rm \dim} \  E_{N+2}^{r-N, N}  ({\rm NL}) $, the number of supersymmetric vacua in the $\zeta \to \infty$ limit according to the Higgs branch analysis in Section 2.3. Since all of the Coulomb branch vacua become Higgs branch vacua in the $\zeta \to \infty$ limit (i.e. the supports of their wavefunctions move towards the $\sigma = 0$ region where the Coulomb branch attaches to the Higgs branch), this disparity cannot be accounted for: something else must be going on.

\paragraph{}
Even so,  the analysis in Section 2.3 is still consistent for certain values of matter sector $R$-charge $\hat r$. For instance, if $\hat r$ lies in the range,
\begin{eqnarray}
\hat r = p_{\rm min} \ , \ ... \  , \ p_{\rm min} + N, \nonumber
\end{eqnarray}
then only $\mathcal H_0^{\hat r} (\sigma < 0 )$ is non-empty. For each element of $\mathcal H_0^{\hat r} (\sigma < 0 )$, we obtain an effective Coulomb branch action for $\sigma \in (-\infty, 0)$, with a single normalisable ground state of total $R$-charge $r = \hat r$. Since $\mathcal H_0^{\hat r} (\sigma < 0 )  \cong E_2^{\hat r, 0 } ({\rm NL}) \cong E_{N+2}^{\hat r, 0} ({\rm NL})$ for $\hat r = p_{\rm min} \ , \ ... \  , \ p_{\rm min} + N$, this is consistent with the counting in the Higgs branch picture.

\paragraph{}
Similarly, if $\hat r$ lies in the range,
\begin{eqnarray}
\hat r = p_{\rm max} + 1 \  , \  ... \ ,\  p_{\rm max} + N+1, \nonumber
\end{eqnarray}
then only $\mathcal H_0^{\hat r} (\sigma > 0 )$ is non-empty, and each element of $\mathcal H_0^{\hat r} (\sigma > 0 )$ corresponds to a single normalisable ground state with $\sigma \in (0, +\infty)$ of total $R$-charge $r = \hat r - 1 $. Since $\mathcal H_0^{\hat r} (\sigma > 0 ) \cong E_2^{\hat r - N - 1, N} ({\rm NL}) \cong E_{N+2}^{\hat r - N - 1, N} ({\rm NL})$ for this range of $\hat r$, this is again consistent with the non-linear model.

\paragraph{}
The novelty arises for $\hat r$ lying in the range,
\begin{eqnarray}
\hat r = p_{\rm min} + N+1 \ , \  ... \ , \  p_{\rm max}. \nonumber
\end{eqnarray}
What is different about $\hat r$ in this range from the perspective of the Coulomb branch picture is that it is possible to find supersymmetric ground states of the matter Hamiltonian both in a background of negative $\sigma$ and in a background of positive $\sigma$. The counting of these matter ground states is related to the non-linear model spectral sequence by
\begin{eqnarray}
\mathcal H_0^{\hat r} (\sigma < 0 ) & \cong & E_2^{\hat r , 0 } ({\rm NL}), \nonumber \\
\mathcal H_0^{\hat r} (\sigma > 0 ) & \cong & E_2^{\hat r - N - 1, N } ({\rm NL}) .\nonumber
\end{eqnarray}
These are precisely the groups acted upon by the differentials
\begin{eqnarray}
 d_{N+1}^{\hat r - N - 1, N}({\rm NL}) : E_2^{\hat r - N-1, N} ({\rm NL}) \to E_2^{\hat r , 0 } ({\rm NL}) , \nonumber
\end{eqnarray}
and the range $\hat r = p_{\rm min} + N+1 \ , \  ... \ , \  p_{\rm max}$ is precisely the range for which this differential can be non-trivial.

\paragraph{}
For $\hat r = p_{\rm min} + N+1 \ , \  ... \ , \  p_{\rm max}$, the ground states of the matter Hamiltonian can be classified into three types:
\begin{itemize}
\item A ground state of the matter Hamiltonian that exists both when $\sigma$ is negative and when $\sigma$ is positive, and depends smoothly on $\sigma $, including when $\sigma$ crosses zero.
\item A ground state of the matter Hamiltonian that exists only when $\sigma$ is negative, and depends smoothly on $\sigma $ for $\sigma < 0 $, but disappears from the spectrum when $\sigma$ is raised to zero.
\item A ground state of the matter Hamiltonian that exists only when $\sigma$ is positive, and depends smoothly on $\sigma $ for $\sigma > 0$, but disappears from the spectrum when $\sigma$ is lowered to zero. 
\end{itemize}

\paragraph{}
Let us define $\mathcal H_0^{\hat r} (\leftrightarrow)$ to be the vector space spanned by the first kind of states, the kind that varies smoothly over the entire range $\sigma \in (-\infty, \infty)$. We can view $\mathcal H_0^{\hat r} (\leftrightarrow)$ both as a subspace of $\mathcal H_0^{\hat r} (\sigma < 0 )$ and as a subspace of $\mathcal H_0^{\hat r} (\sigma > 0 )$.

\paragraph{}
We pick a complementary subspace for $\mathcal H_0^{\hat r}(\leftrightarrow)$ in $\mathcal H_0^{\hat r} (\sigma < 0 )$, spanned by states of the second kind (those that exist for negative background values of $\sigma$ only). We call this complementary subspace $\mathcal H_0^{\hat r} (\leftarrow)$. Thus we have
\begin{eqnarray}
\mathcal H_0^{\hat r} (\sigma < 0 ) & \cong &  \mathcal H_0^{\hat r} (\leftrightarrow ) \oplus \mathcal H_0^{\hat r} (\leftarrow ). \nonumber
\end{eqnarray}
Similarly, we pick a complementary subspace for $\mathcal H_0^{\hat r} (\leftrightarrow)$ in $\mathcal H_0^{\hat r} (\sigma > 0 )$, spanned by states of the third kind, denoting it as $\mathcal H_0^{\hat r} (\rightarrow)$, so that
\begin{eqnarray}
\mathcal H_0^{\hat r} (\sigma > 0 ) & \cong & \mathcal H_0^{\hat r} (\leftrightarrow ) \oplus \mathcal H_0^{\hat r} (\rightarrow ).  \nonumber
\end{eqnarray}

\paragraph{}
Proceeding as in Section 3.3, each element of  $\mathcal H_0^{\hat r} (\leftarrow)$ corresponds to a supersymmetric vacuum of the full theory of  $R$-charge $r = \hat r$, whose wavefunction becomes non-normalisable at $\sigma \to - \infty$ in the limit $ \zeta \to 0$, and approaches the Higgs branch at $\sigma = 0$ in the limit $\zeta \to \infty$.
\paragraph{}
Similarly, each element of $\mathcal H_0^{\hat r} (\rightarrow)$ corresponds to a supersymmetric vacuum of total $R$-charge $r = \hat r   -1 $, whose wavefunction disappears at $\sigma \to + \infty$ in the limit $\zeta \to 0$, and approaches the Higgs branch in the limit $\zeta \to \infty$.

\paragraph{}
It only remains to characterise the effective Coulomb branch actions associated to elements of $\mathcal H_0^{\hat r} (\leftrightarrow)$. The crucial point is that elements of $\mathcal H_0^{\hat r}(\leftrightarrow) $ represent ground states of the matter Hamiltonian that exist for all $\sigma \in (-\infty, + \infty)$, and vary smoothly for this entire range of $\sigma$; therefore, the effective Coulomb branch action corresponding to any element of $\mathcal H_0^{\hat r}(\leftrightarrow)$ is defined for the entire range $-\infty < \sigma < \infty$.

\paragraph{}
As usual, the effective action at lowest order in the derivative expansion is
\begin{eqnarray}
L = \frac 1 {2e^2} (\dot \sigma^2 + i \bar \lambda \dot \lambda + D^2) + h'(\sigma) D - \frac 1 2 h''(\sigma) \bar \lambda \lambda, \nonumber
\end{eqnarray} 
but now, it is understood that $\sigma \in (-\infty, + \infty)$. We analyse how the function $h(\sigma)$ behaves in the various limits. When $|\sigma | \to \infty$, the matter Hamiltonian reduces to a free theory Hamiltonian. So when $\sigma \to - \infty$, the matter ground state reduces to some $\sigma < 0$ free theory ground state, and the potential function $h(\sigma)$ reduces to the potential function for the free theory, which receives its dominant contribution from $\zeta | \sigma |$. Similarly, when $\sigma \to + \infty$, the matter ground state reduces to some $\sigma > 0 $ free theory ground state, and $h(\sigma)$ is dominated by $- \zeta \sigma$.
\begin{eqnarray}
h(\sigma) \sim \begin{cases} + \zeta | \sigma | &  \sigma \to - \infty ;\\  - \zeta \sigma & \sigma \to + \infty. \end{cases} \nonumber
\end{eqnarray}
The matter ground state varies smoothly across $\sigma = 0$, so we expect $h(\sigma)$ to be smooth at $\sigma = 0$ too.

\paragraph{}
The Schr\"odinger equation for this effective Coulomb branch theory has two solutions, 
\begin{eqnarray}
\Psi_1(\sigma) =  e^{-h(\sigma)} | 0 \rangle, \ \ \ \ \  \ \ \  \ \ \  \ \  \ \  \Psi_2(\sigma) =  e^{h(\sigma)} \bar \lambda | 0 \rangle. \nonumber
\end{eqnarray}
But $\Psi_1$ is non-normalisable at $\sigma \to + \infty$ and $\Psi_2$ is non-normalisable at $\sigma \to - \infty$. Higher derivative corrections to the effective theory are unimportant for $| \sigma | \gg e^{3/2}$, so do not affect the failure of these states to be normalisable. The conclusion is that elements of $\mathcal H_0^{\hat r} (\leftrightarrow)$ do not correspond to any supersymmetric vacua in the full theory.

\paragraph{}
We now address the counting of states. On the basis of the information above, we can only say that the number of states of $R$-charge $r$ becoming non-normalisable at the $\sigma \to - \infty $ end of the Coulomb branch as $\zeta \to 0$ is
\begin{eqnarray}
{\rm dim }   E_2^{r, 0} ({\rm NL}) - {\rm dim }  \mathcal H_0^{\hat r = r} (\leftrightarrow),  \nonumber
\end{eqnarray} 
while the number of states of the same $R$-charge becoming non-normalisable at the $\sigma \to + \infty$ end is
\begin{eqnarray}
{\rm dim }   E_2^{r - N, N} ({ \rm NL}) - { \rm dim }  \mathcal H_0^{\hat r = r + 1} (\leftrightarrow). \nonumber
\end{eqnarray}
While this result tightly constrains the possibilities, it does not uniquely determine the number of states of given $R$-charge on each side of the Coulomb branch because we do not know the dimensions of $\mathcal H_0^{\hat r} (\leftrightarrow)$.

\paragraph{}
However, if we \emph{assume} that the number of supersymmetric vacua  in the Coulomb branch picture matches the number in the Higgs branch picture, like we found for complexes of length $\leq N$, then we do obtain a unique possibility for the state counting on each side of the Coulomb branch. It is natural to make this assumption in the context of the models studied in this article because the Higgs branch shrinks to zero size as $\zeta \to 0$, so one would expect that all of the Higgs branch states become visible on the Coulomb branch as $\zeta \to 0$\footnote{There are examples of $\mathcal N = 4$ quiver theories with closed loops where the Higgs branch does not shrink to zero size at the phase boundary, and the agreement between the number of Higgs and Coulomb branch vacua fails: see \cite{purehiggs, purehiggs2, purehiggs3, purehiggs4}.} (though one cannot entirely rule out the possibility that a pair of Higgs branch vacua are lifted when $\zeta$ becomes finite). Nonetheless, in Section 3.6, we will study a concrete example where we can verify explicitly that this assumption holds.

\paragraph{}
As shown in Section 2.3, the total number of Higgs branch vacua of $R$-charge $r$ is
\begin{eqnarray}
{\rm dim} \ E_2^{r, 0} ({\rm NL}) + {\rm dim} \ E_2^{r-N, N} ({\rm NL}) - {\rm rank } \ d_N^{r-N-1,N} ({\rm NL}) - {\rm rank} \ d_N^{r-N,N} ({\rm NL}). \nonumber
\end{eqnarray}
The only way that this can agree with the number of Coulomb branch vacua for all $r$ is if
\begin{eqnarray}
{\rm dim} \ \mathcal H_0^{\hat r} (\leftrightarrow) = {\rm rank} \  d_{N+1}^{\hat r - N - 1, N} ({\rm NL}) \nonumber
\end{eqnarray}
for all $\hat r$. This would imply that the number of vacua of $R$-charge $r$ escaping in the $\sigma \to - \infty$ region is
\begin{eqnarray}
{\rm dim }  \  E_2^{r, 0} ({\rm NL}) - {\rm rank} \ d_{N+1}^{ r - N - 1, N} ({\rm NL}) = {\rm dim } \   E_{N+2}^{r, 0} ({\rm NL}) =  {\rm dim } \   E_{\infty}^{r, 0} ({\rm NL}),  \nonumber
\end{eqnarray} 
and number of vacua of $R$-charge $r$ escaping in the $\sigma \to + \infty$ region is 
\begin{eqnarray}
{\rm dim }   E_2^{r - N, N} ({ \rm NL}) - {\rm rank} \ d_{N+1}^{ r - N, N} ({\rm NL}) = {\rm \dim} \ E_{N+2}^{r-N, N} ({\rm NL}) = {\rm \dim} \ E_{\infty}^{r-N, N} ({\rm NL}). \nonumber
\end{eqnarray}

\subsection{Statement of Main Proposal}
\paragraph{}
Let us summarise:

\paragraph{} We consider the $\mathcal N = 2$ gauged linear sigma model corresponding to the complex of vector bundles,
\begin{eqnarray}
\mathcal F^\bullet = \dots \to \oplus_{j=1}^{D^{p-1}} \mathcal O_{\mathbb P^N} (k_j^{p-1}) \overset{f^{p-1}} {\rightarrow} \oplus_{j=1}^{D^{p}} \mathcal O_{\mathbb P^N} (k_j^{p}) \overset{f^{p}} {\rightarrow} \oplus_{j=1}^{D^{p+1}} \mathcal O_{\mathbb P^N} (k_j^{p+1})  \to \dots \nonumber
\end{eqnarray}
Let $E_s^{p,q}$ be the spectral sequence associated to the bigraded complex,
\begin{eqnarray}
K^{p,q} = \Gamma \left[ \mathcal A^{0, q} \left( \oplus_{j=1}^{D^p} \mathcal O_{\mathbb P^N} (k_j^p) \right) \right] , \ \ \ \  \ \ 
d^{p,q} = \bar \partial,  \ \ \ \ \  {d'}^{p,q} = (-1)^q f^p \nonumber
\end{eqnarray}
For $\zeta \in (0, \infty)$, there are ${\rm dim} \ E_{N+2}^{r,0} + {\rm dim} \ E_{N+2}^{r-N, N}$ supersymmetric vacua of $R$-charge $r$.
\paragraph{}
In the $ \zeta \to \infty $ limit, the vacua are localised on the Higgs branch.

\paragraph{}Assuming that all Higgs branch vacua become visible on the Coulomb branch in the $\zeta \to 0 $ limit, the vacua become non-normalisable at one of the two asymptotic limits of the Coulomb branch:
\begin{itemize}
\item ${\rm dim} \ E_{N+2}^{r,0}$  vacua escape at $\sigma \to - \infty$;
\item ${\rm dim} \ E_{N+2}^{r-N, N}$ vacua escape at $\sigma \to + \infty$.
\end{itemize}

\subsection{Final Example: The Koszul Complex}

\paragraph{}
In this final section, we revisit the three-term Koszul complex on $\mathbb P^1$, example (iv) from Section 2.4, connecting it to the discussion in Section 3.4. This example is particularly instructive because the matter Hamiltonian is quadratic, which enables us to determine exact expressions for the ground states of the matter Hamiltonian, count the number of elements in  $\mathcal H^{\hat r} (\leftarrow)$,  $\mathcal H^{\hat r} (\rightarrow)$ and $\mathcal H^{\hat r} (\leftrightarrow)$ for each $\hat r$, and even determine an explicit expression for the effective vector multiplet theory at lowest order in the derivative expansion.

\paragraph{}
For this example, it is best to work with the Lagrangian written in terms of Fermi multiplets. The gauged linear sigma model has two chiral multiplets $(\phi_i, \psi_i)$ and two Fermi multiplets $\eta_i$, with $i \in \{0, 1 \} $,  all of which have $U(1)$ charge one. The superpotentials are $E_0 = c \phi_0$ and $E_1 = c \phi_1$, where $c$ is a constant parameter with dimensions of energy. 
\begin{eqnarray}
L &= &\frac 1 {2e^2} ( \dot \sigma ^2 + i \bar \lambda \dot \lambda + D^2 ) - \zeta D \nonumber \\
&& + D_t \bar \phi_i D_t \phi_i +  \bar \phi_i (D - \sigma^2 ) \phi_i + i \bar\psi_i D_t \psi_i + \bar \psi_i  \sigma \psi_i - i \bar \phi_i \lambda \psi_i + i \bar\psi_i \bar \lambda \phi_i  \nonumber \\
&& i  \bar\eta_i \dot\eta_i - (u_t + \sigma) \bar \eta_i \eta_i - | c |^2 \bar \phi_i \phi_i - c  \bar \eta_i \psi_i - \bar c \bar \psi_i \eta_i. \nonumber
\end{eqnarray}

\subsection*{Free case: $c = 0$.}

\paragraph{}
If $c = 0$, then integrating out the Fermi multiplets generates a four-dimensional Chan-Paton space. The theory is the direct sum of four decoupled free theories, which can be analysed by the method of \cite{hky} as reviewed in Section 3.1. The associated complex of vector bundles is
\begin{eqnarray}
\mathcal F^\bullet = \mathcal O_{\mathbb P^1}(-2) \overset{0}{\xrightarrow{\hspace*{1.0cm}}} \mathcal O_{\mathbb P^1}(-1)^{\oplus 2}  \overset{0 }{\xrightarrow{\hspace*{1.0cm}}} \underline{\mathcal O_{\mathbb P^1}}.\nonumber
\end{eqnarray}
There are two supersymmetric ground states. The $ \mathcal O_{\mathbb P^2}$ Chan-Paton space supplies one ground state of $R$-charge $0$, localised on the $\sigma < 0 $ part of the Coulomb branch, associated with the lone generator of
\begin{eqnarray}
\mathcal H_0^{\hat r = 0}(\leftarrow)_{c = 0} = E_2^{0,0}({\rm NL})_{c = 0 } = E_{3}^{0,0}({\rm NL})_{c = 0} = \mathbb C. \nonumber
\end{eqnarray}
Additionally, the $\mathcal O_{\mathbb P^2}(-2)[-2]$ Chan-Paton space supplies a further ground state of $R$-charge $-1$, localised on the $\sigma > 0$ branch, associated with the generator of
\begin{eqnarray}
\mathcal H_0^{\hat r = 0}(\rightarrow )_{c = 0} = E_2^{-2,1} ({\rm NL})_{c = 0}  = E_3^{-2,1} ({\rm NL})_{c = 0}  = \mathbb C. \nonumber
\end{eqnarray}

\subsection*{Interacting case: $c \neq 0$.}

\paragraph{}
Let us now turn on a non-zero value for $c$. After integrating out the Fermi multiplets, we are left with interaction terms  corresponding to the sheaf complex,
\begin{eqnarray}
\mathcal F^\bullet = \mathcal O_{\mathbb P^1}(-2) \overset{\left( \begin{array}{c} x_0 \\ x_1 \end{array} \right) }{\xrightarrow{\hspace*{1.4cm}}} \mathcal O_{\mathbb P^1}(-1)^{\oplus 2}  \overset{\left( \begin{array}{cc} -x_1 & x_0  \end{array} \right) }{\xrightarrow{\hspace*{1.4cm}}} \underline{\mathcal O_{\mathbb P^1}}.\nonumber
\end{eqnarray}
As shown in Section 2.4, the non-trivial $s = 2$ terms of the non-linear model spectral sequence are
\begin{eqnarray}
E_2^{0,0}({\rm NL})_{c \neq 0} = E_2^{-2,1} ({\rm NL})_{c \neq 0} = \mathbb C. \nonumber
\end{eqnarray}
However, due to the non-trivial differential $d_2^{-2,1} : E_2^{-2,1} \to E_2^{0,0}$, we have
\begin{eqnarray}
E_3^{0,0}({\rm NL})_{c \neq 0} = E_3^{-2,1} ({\rm NL})_{c \neq 0} = 0. \nonumber
\end{eqnarray}
Therefore, we expect that
\begin{eqnarray}
\mathcal H_0^{\hat r = 0} (\leftarrow)_{c \neq 0 }= \mathcal H_0^{\hat r = 0} (\rightarrow)_{c \neq 0 } = 0, \ \ \ \ \  \ \ \ \mathcal H_0^{\hat r = 0 } (\leftrightarrow)_{c \neq 0} = \mathbb C, \nonumber
\end{eqnarray}
and so we expect to find no supersymmetric ground states.

\paragraph{}
We can see directly how this comes about in the Coulomb branch picture. The matter Hamiltonian in the background $\sigma = {\rm constant}, \lambda =0,  D = 0$ and in the $u_t = 0$ gauge, is
\begin{eqnarray}
H &= & \dot{\bar \phi}_i \dot \phi_i + (\sigma^2 + |c | ^2 ) \bar\phi_i \phi_i \nonumber \\ && - \frac 1 2 \left( \begin{array}{cc}  \bar\psi_i & \bar\eta_i \end{array}\right) \left( \begin{array}{cc} \sigma & \bar c  
\\ c & - \sigma \end{array} \right) \left( \begin{array}{c} \psi_i \nonumber \\ \eta_i  \end{array}\right) + \frac 1 2  \left( \begin{array}{cc}  \psi_i & \eta_i \end{array}\right) \left( \begin{array}{cc} \sigma & c \\ \bar c & - \sigma \end{array} \right) \left( \begin{array}{c} \bar\psi_i \\ \bar\eta_i  \end{array}\right) \nonumber.
\end{eqnarray}
The charge operator is
\begin{eqnarray}
& G = i \bar\phi_i \dot\phi_i - i \dot{\bar\phi}_i \phi_i + \frac 1 2 [\bar\psi_i , \psi_i ] + \frac 1 2 [ \bar\eta_i , \eta_i ] . \nonumber
\end{eqnarray}
To diagonalise $H$ and $G$, we define pairs of creation and annihilation operators,
\begin{eqnarray}
\phi_i = \left( \frac 1 {2\sqrt{\sigma^2 + |c|^2}}\right)^{1/ 2} (\tilde a _i+ \tilde b_i^\dagger ), \ \ \ \ \ \ \dot\phi_i = -i \left( \frac{\sqrt{\sigma^2 + |c |^2 }}{2}\right)^{1/2 } (\tilde a_i - \tilde b_i^\dagger), \nonumber
\end{eqnarray}
and we also define a new set of fermionic operators,
\begin{eqnarray}
\left( \begin{array}{c} \psi_i \\ \eta_i  \end{array} \right) = \frac 1 {\left(2 \left( \sigma^2 + | c |^2 + \sigma \sqrt{\sigma^2 + |c |^2} \right)\right)^{1/2}} \left( \begin{array}{cc} \sigma + \sqrt{\sigma^2 + |c |^2} & - \bar c \\ c & \sigma + \sqrt{\sigma^2 + | c |^2} \end{array} \right) \left( \begin{array}{c}  \chi_i \\ \xi_i \end{array}\right). \nonumber
\end{eqnarray}
As long as $c \neq 0$, the operators $\tilde a_i, \tilde b_i, \chi_i, \xi_i$ are well-defined over the entire  range $\sigma \in (-\infty, \infty)$. (The $c = 0$ case is fundamentally different because $\phi_i$, $\psi_i$ and $\eta_i$ become massless when $\sigma = 0$, so the procedure of integrating them out is invalid at $\sigma = 0$.) Written in the new variables, the Hamiltonian becomes
\begin{eqnarray}
H = \sqrt{\sigma^2 + | c |^2} \left( \tilde a_i^\dagger \tilde a_i + \tilde b_i^\dagger \tilde b_i - \bar\chi_i \chi_i + \bar\xi_i \xi_i + 2 \right), \nonumber
\end{eqnarray}
and the charge operator becomes
\begin{eqnarray}
G = \tilde a_i^\dagger \tilde a_i - \tilde b_i^\dagger \tilde b_i + \bar\chi_i \chi_i + \bar\xi_i \xi_i - 2. \nonumber
\end{eqnarray}
We define a state $|\tilde 0 \rangle $ of $R$-charge $-2$ such that
\begin{eqnarray}
\tilde a_i|\tilde 0 \rangle = \tilde b_i|\tilde 0 \rangle = \chi_i|\tilde 0 \rangle = \xi_i|\tilde 0 \rangle . \nonumber
\end{eqnarray}
It is clear that there is a unique state annihilated by both $H$ and $G$, namely, 
\begin{eqnarray}
\bar\chi_0 \bar\chi_1 | \tilde 0 \rangle.  \nonumber
\end{eqnarray}
Since the operators $\bar\chi_0$ and $\bar\chi_1$ and the state $| \tilde 0 \rangle$ are all well-defined for the entire range $\sigma \in (-\infty, +\infty)$, and vary smoothly with $\sigma$, this supersymmetric ground state counts as an element of $\mathcal H_0^{\hat r = 0 } (\leftrightarrow)_{c \neq 0}$.

\paragraph{}
Let us examine the behaviour of the matter vacuum $\bar\chi_0 \bar\chi_1 | \tilde 0 \rangle$ at each end of the Coulomb branch. For $\sigma \ll - |c |$, we find (up to irrelevant phase factors of $\lambda / |\lambda |)$ that
\begin{eqnarray}
\tilde a_i \sim a_i, \ \ \ \ \tilde b_i \sim b_i, \ \ \ \ \chi_i \sim \eta_i, \ \ \ \ \xi_i \sim \psi_i , \nonumber
\end{eqnarray}
so as $\sigma \to - \infty$, the matter vacuum $\bar\chi_0 \bar\chi_1 | \tilde 0 \rangle$  reduces to the matter vacuum of the $c = 0$ free theory associated with the $\mathcal O_{\mathbb P^2}$ Chan-Paton state,
\begin{eqnarray}
\bar\chi_0 \bar\chi_1 | \tilde 0 \rangle \sim \bar\eta_0 \bar\eta_1 |0 \rangle =  |\mathcal O_{\mathbb P^2} \rangle. \nonumber
\end{eqnarray}
For $\sigma \gg + |c |$, we have
\begin{eqnarray}
\tilde a_i \sim a_i , \ \ \ \ \tilde b_i \sim b_i , \ \ \ \ \chi_i \sim \psi_i , \ \ \ \ \xi_i \sim \eta_i, \nonumber
\end{eqnarray}
so as $\sigma \to + \infty$, the matter vacuum $\bar\chi_0 \bar\chi_1 | \tilde 0 \rangle$ reduces to the free theory vacuum associated with the $\mathcal O_{\mathbb P^2}(-2) $ Chan-Paton state,
\begin{eqnarray}
\bar\chi_0 \bar\chi_1 | \tilde 0 \rangle \sim \bar\psi_0 \bar\psi_1 |0 \rangle = \bar\psi_0 \bar\psi_1 |\mathcal O_{\mathbb P^2} (-2)\rangle. \nonumber
\end{eqnarray}

\paragraph{}
Next, in order to determine the effective Coulomb branch action associated to $\bar\chi_0 \bar\chi_1 | \tilde 0 \rangle$, we examine how this matter ground state responds to fluctuations in $D$. If we turn on a non-zero value for $D$, then, in order to diagonalise the matter Hamiltonian, we must replace
\begin{eqnarray}
\sqrt{\sigma^2 + | c |^2 } \  \mapsto \ \sqrt{\sigma^2 - D + | c |^2 } \approx \sqrt{\sigma^2 + |c |^2} - \frac D {2 \sqrt{\sigma^2 + | c |^2}},  \nonumber
\end{eqnarray}
in the definitions of $\tilde a_i$ and $\tilde b_i$, while retaining the old expressions for $\chi_i$ and $\xi_i$. We make the same replacement in front of the $\tilde a_i^\dagger \tilde a_i + \tilde b_i^\dagger \tilde b_i + 2$ term in the Hamiltonian. So the shift in the energy of the state $\bar\chi_0 \bar\chi_1 | \tilde 0 \rangle $ is
\begin{eqnarray}
\Delta H = - \frac{D}{\sqrt{\sigma^2 + |c |^2 }} .\nonumber
\end{eqnarray}
To reproduce this shift in energy, the effective Lagrangian for the vector multiplet, at lowest order in the derivative expansion, must be
\begin{eqnarray}
L = \frac 1 {2e^2} (\dot \sigma^2 + i \bar \lambda \dot \lambda + D^2) + h'(\sigma) D - \frac 1 2 h''(\sigma) \bar \lambda \lambda \nonumber
\end{eqnarray}
with $\sigma $ spanning the entire range of the real line, and
\begin{eqnarray}
h(\sigma) = \sinh^{-1} \left( \frac \sigma { | c | } \right)- \zeta \sigma. \nonumber
\end{eqnarray}
Solving the Schr\"odinger equation for this effective Lagrangian gives two states of zero energy,
\begin{eqnarray}
\Psi_1 (\sigma) &  = & \frac 1 2  \left( - \sigma + \sqrt{\sigma^2 + | c |^2} \right)e^{+ \zeta \sigma} | 0 \rangle, \nonumber \\
\Psi_2 (\sigma) & = &  \frac 1 2 \left( + \sigma + \sqrt{\sigma^2 + | c |^2 } \right) e^{-\zeta \sigma} \bar \lambda  | 0 \rangle. \nonumber
\end{eqnarray}
But neither state is normalisable at both $\sigma \to - \infty$ and $\sigma \to +\infty$. Hence there are no supersymmetric vacua.

\paragraph{}
It is instructive to examine once more what happens in the free theory limit. When $c \to 0$, the potential function $h(\sigma)$ diverges at $\sigma = 0$. $\Psi_1(\sigma)$ becomes a normalisable wavefunction supported entirely on $\sigma \in (-\infty, 0)$, and corresponds to the generator of $\mathcal H_0^{\hat r = 0} (\leftarrow)_{c = 0}$.  Meanwhile, $\Psi_2(\sigma)$ becomes a normalisable wavefunction supported on the other side of the Coulomb branch, $\sigma \in (0, + \infty)$, and corresponds to the generator of $\mathcal H_0^{\hat r = 0} (\rightarrow)_{c = 0}$.

\paragraph{}
Thus, the physical interpretation of the non-trivial differential
\begin{eqnarray}
d_2^{-2,1}: E_2^{-2,1}({\rm NL})_{c \neq 0} \to E_2^{0,0} ({\rm NL})_{c \neq 0}. \nonumber
\end{eqnarray}
 is that it smoothens out the singularity in the potential function $h(\sigma)$ at $\sigma = 0$, allowing the normalisable $\mathcal H_0^{\hat r = 0} (\leftarrow)_{c = 0}$ and $\mathcal H_0^{\hat r = 0} (\rightarrow)_{c = 0}$ states to tunnel across the $\sigma = 0$ region and merge into the pair of non-normalisable states associated with the lone generator of  $\mathcal H_0^{\hat r = 0 } (\leftrightarrow)_{c \neq 0} \cong \mathbb C$. 

\section{Concluding Remarks}

\paragraph{}
This work fits into a growing body of literature on wall-crossing phenemena in supersymmetric quantum mechanics (see e.g. \cite{ sungjay}, \cite{ sen} - \cite{morewall5} for example). What is appealing about the perspective taken in the present work is that it allows us to visualise wall-crossing events in terms of states becoming non-normalisable in certain regions of the Coulomb branch, along the lines of Denef \cite{denef}, even in theories with  potentials.

\paragraph{}
Throughout this work, we  focused exclusively on models with a $U(1)$ gauge field and chiral fields of charge $+1$. This class of models includes many interesting ones, notably the GLSMs for complete intersections in projective space, but one may well ask whether the techniques are applicable to more general models. It is simple to modify the analysis to accommodate models with a $U(1)$ gauge field and chiral fields of positive but unequal charge: the theories are then associated with complexes of line bundles on weighted projective space, whose global sections are elements of given grading in weighted polynomial rings, and the same kind of analysis goes through. However, it is unclear whether the analysis generalises to $U(1)$ models with both positive and negative chiral fields, which have non-compact target spaces. The difficulty here is that the analysis in Section 3.3 requires that we choose one of the two supercharges to be the differential in a spectral sequence, depending on the signs of the charges of the fields, and with both positive and negative chiral fields present, neither choice of supercharge simplifies the calculation in an obvious way. It would also be interesting to understand whether certain aspects of this analysis can be applied to theories with higher rank or non-abelian gauge groups, as this could provide physical intuition for the combinatorics of certain cases of the Borel-Weil-Bott theorem.

\paragraph{}
Finally, one may ask whether the present techniques are relevant to two-dimensional models with supersymmetric boundaries. A complication that arises with two-dimensional theories is that the Coulomb branch variable is complex rather than real, and the boundary of the Coulomb branch at infinity is a circle rather than a pair of points. Therefore, it appears at first glance that analysing the Coulomb branch of these two-dimensional theories requires a fundamentally different approach. Nonetheless, this question is interesting from a mathematical perspective, because counting vacua in two-dimensional theories gives full information about the morphisms in the derived category of the target space, rather than merely the hypercohomology groups. We leave this fascinating problem for future work.

\section*{Acknowledgements}
The author would like to thank Ed Segal for a vital conversation. The author is supported by Gonville and Caius College and the ERC Grant agreement STG 279943.

\end{document}